# Chiral Self-Assembly of Cellulose Nanocrystals is Driven by Crystallite Bundles

**Authors**

Thomas G. Parton, Richard M. Parker, Gea T. van de Kerkhof, Aurimas Narkevicius, Johannes S. Haataja, Bruno Frka-Petesic and Silvia Vignolini*

**Affiliation**

[1]Yusuf Hamied Department of Chemistry, University of Cambridge, Lensfield Road, Cambridge CB2 1EW, United Kingdom

*Corresponding author. sv319@cam.ac.uk

**Abstract**

The transfer of chirality across length-scales is an intriguing and universal natural phenomenon. However, connecting the properties of individual building blocks to the emergent features of their resulting large-scale structure remains a challenge. In this work, we investigate the origins of mesophase chirality in cellulose nanocrystal suspensions, whose self-assembly into chiral photonic films has attracted significant interest. By correlating the ensemble behaviour in suspensions and films with a quantitative morphological analysis of the individual nanoparticles, we reveal an inverse relationship between the cholesteric pitch and the abundance of laterally-bound composite particles. These 'bundles' thus act as colloidal chiral dopants, analogous to those used in molecular liquid crystals, providing the missing link in the hierarchical transfer of chirality from the molecular to the colloidal scale.

## Introduction

The chiral self-assembly of nanoscale building blocks is a universal phenomenon that demonstrates the emergence of large-scale structure from the properties of individual sub-units, and offers a way to produce functional materials with bespoke properties[1,2]. In many self-organising colloidal systems such as amyloid fibrils and DNA origami filaments, the emergence of chirality in the mesophase can be directly correlated to the well-defined chiral morphology of individual assembling elements, or their bending properties[3–6]. In general, however, directly tracking the transfer of chirality across length-scales remains a significant challenge, both for model colloids such as rod-like viruses and in systems of practical relevance[7].

Cellulose nanocrystals (CNCs) are bio-sourced nanoparticles that spontaneously self-assemble in colloidal suspension to form a left-handed chiral nematic (cholesteric) phase (**Fig. 1a**).[8–10] While their predisposition to self-assemble has been exploited as a chiral scaffold[11] or to produce sustainable photonic colourants[12,13], the origin on their mesophase chirality remains a highly debated topic. Individual cellulose Iβ crystallites are elongated and have a propensity to twist about their long axis[14–16], a property ultimately arising from the molecular chirality of the β-1,4-D-glucose repeating unit[17–19], but it is unclear how (or if) this subtle twist is sufficient to induce the chirality of the mesophase[20]. Moreover, the CNC population is highly polydisperse in particle size and shape[21], which complicates attempts to attribute the chirality of the mesophase to specific features of individual particles.

By applying ultrasonication to controllably tune the size and shape of the CNCs and performing detailed nanoparticle shape metrology, here we establish quantitative relationships between the colloidal liquid crystalline behaviour and the morphological distribution of individual particles. This analysis reveals that a sub-population of composite particles, which we refer to as ‘bundles’, acts as colloidal chiral dopants, analogous to those used in molecular liquid crystals to transform a nematic phase into a cholesteric phase of a desired chiral strength[22]. As such, this ubiquitous but often ignored component of CNC suspensions is essential for the transfer of chirality across length-scales.

## Results

The strength of the chiral interactions in CNC suspensions can be accessed by measuring the chiral nematic pitch (defined as the period of one full rotation of the helicoidal structure), with weaker interactions corresponding to a larger pitch. While the pitch in suspension is typically on the order of 5-50 μm, compression of the structure upon evaporation of the solvent results in a periodicity of 250-500 nm in dried films, giving rise to structural colour in the visible range[9]. The visual appearance of photonic CNC films is commonly tuned by treating the initial suspension with high-intensity ultrasound (hereafter referred to as ‘sonication’), which causes a red-shift in the colour of the films by increasing the chiral nematic pitch in suspension[23]. However, the mechanism by which sonication weakens the chiral interactions between individual CNCs is unclear.

To explore the effect of sonication on CNC self-assembly, we produced an initial CNC suspension by sulphuric acid hydrolysis of cotton (Methods) and applied sonication over a wide range of doses ($u_S$ = 0-12 kJ/mL) to produce photonic films with colour across the visible range (**Fig. 1b**). Importantly, the sonication dose is expressed as caloric energy per suspension volume (J/mL), as we observed that consistent results could only be obtained using this unit (see Supplementary Information section S2). The pitch, as measured by cross-sectional scanning electron microscopy (SEM), was found to increase with sonication dose (**Fig. 1c**, Supplementary Figure 5). For photonic films with visible structural colour, the pitch values from cross-sectional SEM are consistent with the peak wavelengths from optical spectroscopy (Supplementary Figure 6).

Identifying the origin of the pitch increase is challenging due to the numerous potential effects of sonication on individual particles, both physical (e.g. size, aspect ratio) and chemical (e.g. surface charge)[23]. By careful control of the suspension conditions after sonication (e.g. ionic strength) and characterisation of suspension properties (e.g. surface charge, pH), we concluded that the chiral interaction between CNCs was fundamentally entropic, and the observed increase in pitch could only arise from sonication-induced changes in particle size and shape (see Supplementary Information section S3 for further discussion). This conclusion is consistent with reports of chiral nematic phase formation in refractive index-matching apolar solvents, such as toluene, where electrostatic and van der Waals interactions are suppressed[24]. However, the overall trends in ensemble size measurements such as hydrodynamic diameter or particle cross-section (Supplementary Figure S15) are not consistent with the observed pitch increase, suggesting that the crucial morphological changes are more subtle and require more detailed characterisation.

The morphology of individual particles can be observed using transmission electron microscopy (TEM), as exemplified in **Fig. 1d** and **Fig. 1e**. The preparation of TEM grids was optimised to remove physical overlap of distinct particles (see Methods), and we therefore define a CNC particle, in the context of this study, as any continuous nano-object observed in TEM images (see Supplementary Information section 1.10 for further discussion). While all CNC suspensions exhibit considerable variation in particle size and shape, there are clear morphological trends with sonication. At low dose, the CNC population contains a large number of irregular-shaped particles that are composed of multiple elongated sub-units bound perpendicular to their long axis (**Fig. 1d**). The dimensions of the sub-units are consistent with those expected for individual cellulose crystallites, which indicates that the CNC population is a mixture of isolated elementary crystallites and multi-crystallite composite particles, in agreement with previous reports for CNCs derived from wood pulp and cotton.[25,26] At higher sonication doses, the fragmentation of composite particles leads to a higher proportion of isolated cellulose crystallites (**Fig. 1e**).

The sonicated-induced release of individual cellulose crystallites is noteworthy, as their right-handed axial twist is often cited as the likely origin of mesophase chirality in CNC suspensions,[16,27] and is supported by simulations of monodisperse hard polyhedral particles with a fixed axial twist[28]. However, if the mesophase chirality arose directly from the twisted morphology of individual cellulose crystallites, the pitch would be expected to remain unchanged or even decrease as the crystallites are liberated from achiral composite particles by sonication. In practice the opposite trend is observed, suggesting that the crystallite twist is not, by itself, sufficient to explain the mesophase chirality.

This conclusion led us to consider the possible role of composite particles in CNC self-assembly. Historically, these particles have often been overlooked when manually extracting morphological properties from TEM images, on the assumption that they are an artefact of sample preparation. Having first verified that multi-crystallite particles are a native component of the particle population using cryogenic TEM images of dilute CNC suspensions (Supplementary Figure 7), we sought morphological properties to characterise particle size and shape. Individual cellulose crystallites have a regular, highly elongated morphology, making it appropriate to represent their outlines as simple model shapes such as rectangles or ellipses. However, to accurately capture the complex, irregular shapes of many CNCs (**Fig. 2a**), such approximations are insufficient. For instance, fitting the particle shape to a bounding rectangle of dimensions $L_b$ and $W_b$ over-estimates the projected particle area ($A \leq L_b W_b$) and under-estimates the aspect ratio (**Fig. 2b,c**), making it difficult to correlate subtle changes in particle morphology with ensemble behaviour.

We therefore developed an in-depth morphological analysis protocol and applied it to CNC outlines obtained from TEM images (Materials and Methods), to quantify the evolution in particle size and shape with sonication dose. This detailed nanoparticle metrology allows us to measure more appropriate size properties, such as the area-equivalent (AE) width, $W_{AE} = A/L_b$ which better represents the average width of elongated but irregular particles like those found in CNC suspensions.

The sonication-induced fragmentation of CNCs can be understood by examining the change in the key morphological properties. This is exemplified by **Fig. 2d-e**, which show the size distributions of $L_b$ and $W_{AE}$ for CNC suspensions exposed to a range of sonication doses. The general trend is a decrease in particle dimensions with sonication dose, which we also observed for other common size properties such as area or perimeter (Supplementary Figure 8). Notably, while the mean box length (**Fig. 2d**) decreases gradually with sonication dose, the mean AE width (**Fig. 2e**) converges towards a limiting value of 8 nm, comparable to that expected for a single cellulose crystallite[25]. Furthermore, the narrowing of the length and width distributions in **Fig. 2d-e** with sonication dose indicates a decrease in the coefficient of variation (*i.e.* standard deviation over mean, a measure of relative polydispersity).

Alongside size properties, this analysis also allows us to quantify size-invariant shape properties such as particle rectangularity, defined as

$$\mathcal{R} = A/L_{\mathrm{b}}W_{\mathrm{b}} = W_{\mathrm{AE}}/W_{\mathrm{b}}. \tag{1}$$

The rectangularity ranges from 0 to 1, with $\mathcal{R} = 1$ for a particle with an ideal rectangular profile. The increase in mean particle rectangularity (**Fig. 2f**) provides robust statistical evidence of the simplification of particle shape with sonication dose, which is consistent with the qualitative observation of fragmentation leading to individual parallelepipedic crystallites. Further evidence of this trend is readily obtained from other shape properties, such as particle convexity and solidity (Supplementary Figure 9).

Another fundamental shape properties for elongated particles is the aspect ratio (i.e. the ratio of longer and shorter particle dimensions). However, accurately measuring the aspect ratio of CNCs is non-trivial due to their irregular shapes and their non-circular cross-sections, which are not visible in TEM images. For individual cellulose crystallites, the cross-section perpendicular to the long axis is known to be approximately rectangular (**Fig. 2c**), with a width and thickness of around 6-8 nm for crystallites derived from cotton[25,29]. More generally, the 3D morphology of CNC particles can be probed by atomic force microscopy (AFM) or small angle scattering, with previous studies indicating that many CNCs are composed of laterally-bound crystallites into "bundled" or "raft-like" structures[25,30,31]. By combining TEM and AFM analysis, we estimated the mean thickness $\langle T \rangle$ of each particle observed in TEM (Supplementary Information, section S8), which allows us to predict the effective 3D aspect ratio $\alpha_{3D} = L_b/\sqrt{W_{\mathrm{AE}}\langle T \rangle}$, as opposed to the apparent 2D aspect ratio obtained from TEM (i.e. $\alpha_{2D} = L_b/W_b$). The relevance of $\alpha_{3D}$ is demonstrated by its ability to accurately estimate ensemble size properties such as the average hydrodynamic diameter and particle cross-section from the morphology of individual particles (Supplementary Figure 15). This 3D morphological analysis also gave access to a variety of other important shape properties such as the aspect ratio of the particle cross-section ($\alpha_{\mathrm{XS,AE}} = W_{\mathrm{AE}}/\langle T \rangle$) and the 3D isoperimetric quotient (Supplementary Figure 10).

The 3D aspect ratio decreased substantially with sonication dose (**Fig. 2g**). Such a decrease is notable because of its predicted effect on the lyotropic phase behaviour of CNC suspensions. In the limit of infinite dilution, CNC suspensions are isotropic (I), with no correlation in particle orientations. Above a critical CNC volume fraction $\phi_0$, the chiral nematic (N*) phase is first observed in coexistence with the isotropic phase. The relative proportion of the N* phase grows with concentration until the suspension becomes fully chiral nematic at a second critical volume fraction $\phi_1$. Notably, the critical concentrations ($\phi_0, \phi_1$) are predicted to be inversely proportional to the particle aspect ratio[32]. At higher concentrations, equilibration can be inhibited by the onset of kinetic arrest (KA) with the mechanism of KA (whether glass transition or gelation) dependent on the colloidal conditions[9,33].

To investigate the correlations between CNC phase behaviour and particle morphology, we prepared sealed capillaries containing suspensions at a range of concentrations and sonication doses and allowed them to reach equilibrium (**Fig. 3a**). As shown in **Fig. 3c**, we found that the onset of the biphasic concentration range ($\phi_0$) was not strongly affected by sonication, while the onset of the fully chiral nematic phase ($\phi_1$) was delayed. These trends are broadly consistent with the decrease in mean 3D aspect ratio with sonication dose (Supplementary Information section S10). At the highest dose ($u_S$ = 12 kJ/mL), sonication advanced the onset of kinetic arrest, with the low ionic strength of the suspensions and microscale appearance being consistent with a repulsive colloidal glass[33]. This observation may be related to the greater effective volume fraction in suspension for these highly fragmented particles[34].

Importantly, the capillaries also enable measurement of the chiral nematic pitch in suspension, which appears as a periodic "fingerprint pattern" visible by polarised optical microscopy (**Fig. 3b**). At concentrations below the onset of the kinetic arrest, the equilibrium pitch was found to be inversely proportional to the CNC volume fraction at all sonication doses (**Fig. 3d**). For samples at comparable CNC concentrations, we found that sonication universally led to a larger pitch, indicating that the red-shift observed in dried films originates from a larger pitch in the suspension, and not from variation in the compression of the chiral nematic structure upon drying. These observations can be formulated as a linear relationship between the chiral nematic wavevector, $q = 2\pi/P$, and the CNC volume fraction, $\phi$:

$$q = \kappa\phi, \quad (2)$$

where $\kappa(u_S)$ is a dose-dependent property that captures the strength of the chiral interactions in the suspension. Expressing the mesophase chirality in terms of this chiral strength $\kappa$, instead of the pitch, enables direct comparison between values at a wide range of concentrations. We observe that the chiral strength decreases with sonication dose according to an empirical logarithmic fit (**Fig. 3e**).

Remarkably, the linear scaling in **Eq. 2** is reminiscent of the behaviour of chiral doping in molecular liquid crystals[22]. For a nematic liquid crystal doped by chiral molecules at a volume fraction $\phi_d$, the wavevector of the resulting chiral nematic phase is

$$q = 4\pi\beta_V\phi_d, \quad (3)$$

where $\beta_V$ is the helical twisting power (HTP) of the chiral dopants. The scaling behaviour in CNC suspensions thus suggests that only a small proportion of the CNC population actively contribute to the mesophase chirality. We hypothesised that a specific sub-population of the composite CNC particles act as chiral dopants, while isolated elementary crystallites are functionally achiral and tend to form a nematic (i.e. non-cholesteric) phase. The decrease in chiral strength with sonication dose can then be attributed to the fragmentation of these composite particles into elementary crystallites.

To validate this hypothesis, we mixed two CNC suspensions of different sonication doses at constant CNC concentration. We chose 'low-dose' (772 J/mL) and 'high-dose' (3087 J/mL) samples with substantially different pitch values at the chosen concentration ($\phi$ = 7.9 vol%) but comparable mean particle length and AE width. This concentration was chosen to ensure both samples were fully chiral nematic, ruling out any possible complications due to fractionation within a biphasic suspension[34,35]. The equilibrium pitch values for mixtures of low and high dose samples (**Fig. 3f**) are consistent with the chiral dopant hypothesis, which predicts a harmonic pitch average for the mixtures (*i.e.* a linear mixing in terms of the corresponding wavevectors $q$), due to variation of the abundance of dopants according to **Eq. 3**. Notably, the trend in **Fig. 3f** is not consistent with pitch variation arising from bulk suspension properties such as ionic strength, where the resulting pitch would be a linear interpolation of the pitches of the two original samples, as previously proposed[23].

The evolution of the pitch in CNC mixtures reported in **Fig. 3f** motivated us to re-examine the CNC morphological analysis for evidence of a sub-population acting as chiral dopants. The morphological analysis presents a complex picture of the evolution of the particle population with increasing sonication dose. For instance, in **Fig. 2f** a low rectangularity ($\mathcal{R}$) tail was observed at low dose, which disappeared at intermediate doses but re-emerged at high doses. Direct inspection of TEM images revealed that low-dose suspensions ($u_S \leq 48$ J/mL) had a considerable number of large, irregularly-shaped particles with low rectangularity (*i.e.* shapes that were poorly approximated by a bounding rectangle). The proportion of these irregular particles decreased rapidly as the sonication dose increased. In contrast, much higher doses ($u_S \geq 772$ J/mL) led to the occurrence of particles that appeared to be elementary crystallites distorted by sharp kinks. Although these kinked particles also showed low rectangularity, they were markedly thinner than the low-rectangularity particles observed at lower doses. Furthermore, their morphology was similar to previous reports for CNCs extracted from tunicate or for CNCs produced by sonicating cellulose nanofibers[25,36], leading us to conclude that the kinked particles were elementary crystallites that had been mechanically damaged by extensive sonication.

A plausible model of a chiral composite particles is a 'twisted raft' or 'propeller' configuration, where crystallites are bound together perpendicular to their long axes, with a twist about the binding axis arising from the axial twist on the crystallite sub-units. A recent simulation study has demonstrated that an ensemble of hard homochiral 'twisted raft' particles will form a chiral nematic phase, with pitch values comparable to CNCs suspended in apolar solvents.[37] More generally, among the diversity of particle shapes observed in CNC samples, we anticipate an enantiomeric excess of composite particles with a right-handed twisted morphology with the symmetry-breaking arising from the weak right-handed twist on individual crystallites. Previous studies applying cryogenic electron tomography to individual CNCs have observed chiral composite particles, but the practical limitations of this technique have made it difficult to establish a net chiral bias[38]. It is difficult to discern the 3D chiral morphology of CNCs from the individual 2D projections provided by TEM images, and the drying of the sample onto the TEM grid leads to distortion of the chiral structure.[16] However, the kind of structures necessary for chiral doping should be distinguishable from disordered aggregates and individual cellulose crystallite (whether native or kinked) by their morphological properties. In this regard, drying the sample onto a TEM (or AFM) grid has the advantage of flattening particles onto the grid, allowing their dimensions to be measured without relying on estimation of the projected shape[25].

We therefore classified CNCs into different sub-populations according to their AE width and rectangularity, with threshold boundaries defined at $W_{AE}$ = 17 nm and $\mathcal{R}$ = 0.40. The width threshold was chosen so that particles above this threshold are unquestionably composite particles (i.e. $W_{AE} > 2\,W_c$, where $W_c$ is the average width of a single crystallite), while the rectangularity threshold was chosen based on the shape of the distributions in **Fig. 2f**. This filtering by two morphological properties leads to four particle classes. For convenience, each class is named according to the morphology of an archetypal particle: (A) Aggregates ($W_{AE}$ high, $\mathcal{R}$ low), (B) Bundles ($W_{AE}$ high, $\mathcal{R}$ high), (C) Crystallites ($W_{AE}$ low, $\mathcal{R}$ high), and (D) Distorted (kinked) crystallites ($W_{AE}$ low, $\mathcal{R}$ low). The labelling of the classes does not perfectly describe every particle in each sub-population (especially for the Crystallite sub-population), but captures the essential features. The wide array of available morphological properties offers many alternative ways to divide CNCs into multiple classes (e.g. based on the aspect ratio of the particle cross-section, Supplementary Figure 10). The classification above was chosen as a simple but effective example that relies purely on properties directly observed from TEM images, without relying on estimation of the particle thickness or other 3D properties not accessible by TEM.

**Fig. 4a** displays the population distributions of CNC particles at selected sonication doses. The overall trend with sonication is a migration of the particle distribution counter-clockwise in the $W_{AE} - \mathcal{R}$ parameter space, from Aggregates (bottom right) through Bundles (top right) and Crystallites (top left) to Distorted crystallites (bottom left). This evolution in particle morphology is represented in **Fig. 4b**, which shows typical particles in each sub-population. We then quantified the relative volume fraction ($\mathcal{V}_A$, $\mathcal{V}_B$, *etc.*) of each particle sub-population (Supplementary Information, Section S11), which allowed us to track the evolution of the CNC population with sonication, as shown in **Fig. 4c.** For completeness, the evolution of the relative number fractions of each sub-population can be found in Supplementary Figure 17, and histograms for various morphological properties of the sub-populations can be found in Supplementary Figures 18-20.

The trends in **Fig. 4c** match the qualitative observations of **Fig. 4a**, and are consistent with the ensemble size measurements such as suspension turbidity or hydrodynamic diameter (Supplementary Figure 15). While ensemble measurements are only an indirect measure of particle dimensions, any morphological differences observed between samples are highly robust. In this case, a marked decrease in turbidity (measured by UV-vis spectroscopy) and hydrodynamic diameter (measured by dynamic light scattering) was observed, with both displaying a rapid initial decrease at low sonication doses and then a more gradual decrease at higher doses (Supplementary Figure 15). In terms of our particle classification, these observations confirm that the Aggregate and Bundle sub-populations are initially dominant and that, upon sonication, Aggregates are readily converted into smaller structures while the Bundles are more persistent and require higher doses to be broken down into individual elementary crystallites.

Having validated the evolution of particle morphology by comparing to ensemble measurements, we considered whether a specific sub-population act as chiral dopants. The dopant concentration can be related to the total CNC concentration by defining a relative dopant volume fraction $\chi$, which can be obtained by combining **Eq. 2** and **Eq. 3**:

$$\chi = \phi_d/\phi \, . \tag{4}$$

By comparing sub-population trends in **Fig. 4c** to the corresponding chiral strength values in **Fig. 3e**, we identified the Bundle sub-population as potential chiral dopants (*i.e.* $\chi = \mathcal{V}_B$), as expected by the 'twisted raft' model. We discounted the Aggregates as potential chiral dopants because this sub-population is absent at moderate to high sonication doses, and their morphology is highly irregular. Chirality transfer is expected to be stronger between particles with compatible morphological properties (e.g. isoperimetric quotient), and in this respect the Crystallites are much closer to the Bundles than to the Aggregates (see histograms in Supplementary Information Figure 20).[39] Furthermore, chirality transfer requires a non-circular particle cross-section, and the Bundles are found to have a substantially larger cross-section aspect ratio than the Crystallite and Distorted crystallite sub-populations (Supplementary Figure 19).

The chiral strength $\kappa$, extrapolated from the measurements of the volume fraction and the pitch, shows a clear positive trend with dopant abundance, as shown in **Fig. 4d**. It also increases linearly with $\chi$ at low Bundle abundance, in accordance with **Eq. 4**, implying a constant HTP at high sonication dose. Notably, the chiral nematic wavevector $q$ tends to zero at high sonication, supporting the conclusion that the crystallites are functionally achiral.

The limiting behaviour of **Fig. 4d** gives an estimate of the HTP per bundle volume fraction as $\beta_V$ = – 4 $\mu m^{-1}$, where the negative sign is used to indicate left-handed helicoidal ordering. This value is comparable to those obtained for cholesteric colloidal systems of DNA origami filaments and amyloid fibrils, where the estimated helical twisting powers were $|\beta_V| \approx 2$ and $|\beta_V| \approx 20$ $\mu m^{-1}$ respectively[3,4]. While the chiral dopant model is expected to be universally applicable to the mesophase behaviour of any CNC suspension, we expect the fundamental properties of the Bundle sub-population (i.e. helical twisting power and relative abundance) to vary slightly with cellulose source and extraction method. Interestingly, by considering the HTP per mass fraction ($\beta_w \approx -2.5$ $\mu m^{-1}$), a comparison can be made with previous reports of functionalised CNCs doping a molecular nematic mesophase ($\beta_w \approx -50$ $\mu m^{-1}$). The more effective inducement of chirality in the latter case can be attributed to the much smaller mesogen size and continuous short-range contact between the mesogens and chiral dopants.[40]

## Discussion

These results prompt a re-examination of the existing literature on the chiral self-assembly of CNCs. First, it has been shown that fractionation of CNC suspensions by phase separation leads to an accumulation of longer particles in the anisotropic phase, with a positive correlation between mean particle length and helical twisting power[12,34,35]. Here, we observed that the Bundle sub-population has greater mean particle length and 3D aspect ratio than the Crystallites (Supplementary Figure 18), and that the helical twisting power is proportional to the relative volume fraction of this sub-population. We therefore interpret the observed fractionation as an indication that Bundles preferentially enter the anisotropic phase. Second, our model is consistent with the observed decrease in pitch when increasing the ionic strength of CNC suspensions[41], as greater screening of the electrostatic repulsion between particles allows them to come into closer proximity, where the entropic (i.e. short-range) chiral interaction can take effect. Finally, composite CNCs acting as chiral dopants offer a convincing and self-consistent explanation for the pitch increase upon sonication, in contrast to the previously proposed model of Beck *et al.*,[23] where the pitch increase was attributed to the release of tightly-bound ions and an increase in the effective thickness of the CNC surface layer (Supplementary Information, section S3). Importantly, the increase in chiral nematic pitch due to sonication has been observed for suspensions of polysaccharide nanocrystals from numerous sources, including CNCs from cotton[23] and wood pulp[13], as well as chitin nanocrystals,[42] suggesting that the chiral dopant model can be applied to understanding the chirality of these systems[43]. Despite the strong evidence of the chiral dopant model, detailed quantitative analysis of 3D CNC morphology sufficient to demonstrate an enantiomeric excess of chiral particles is still lacking. However, it is important to note that the chiral dopant model is widely applicable, regardless of the specific dopant shape, and is not limited to the specific composite particle shape (the 'twisted raft') assumed in this work.

The role of CNC bundles as chiral enhancers suggests a new paradigm for colloidal self-assembly, analogous to chiral dopants in molecular liquid crystals. It is difficult to manufacture microscale particles with a consistent chiral morphology, often requiring intensive top-down methods or expensive starting materials. Our results suggest that large-scale complex self-assembly could instead be achieved by adding a small fraction of precisely designed chiral seed particles into an achiral bulk phase produced by simpler methods.

The detailed nanoparticle metrology used in this work opens up a new avenue of enquiry for research on anisotropic bio-sourced nanomaterials. Beyond the role of chirality, nanocellulose formulations prepared from different sources and extraction methods exhibit considerable variation in performance[44,45], which could be systematically analysed and correlated with the prevalence of specific morphological features. More generally, standardising the characterisation of nanoparticle morphology is a major outstanding challenge for nanomaterials research [46,47]. The particle analysis developed for this study, which is widely transposable to other systems, may help address this challenge.

## Methods

### Preparation of CNC suspensions

Cellulose nanocrystals were obtained by acid hydrolysis of Whatman No 1 cellulose filter paper (60 g) with sulphuric acid (64 wt%) at 64 °C under high mechanical stirring. The reaction was quenched after 30 min by dilution of the acid with deionised ice and water and immersion of the reaction vessel in an ice bath. Soluble cellulose residues and excess acid were removed by three rounds of centrifugation at 20,000 g (30, 20, 20 min respectively) with the pellet redispersed in deionised water after each round. Excess ions were removed by dialysis against deionised water using MWCO 12-14kDa membranes. This purification resulted in a 2.43 wt% CNC suspension, which was used as the starting batch for further treatment.

### Determination of CNC mass and volume fraction

The concentration of CNC suspensions was measured by a thermogravimetric method. Vials containing CNC suspensions were weighed before and after drying (>24 hours at 60 °C) to determine the CNC mass fraction $\mu_{CNC}$. The CNC volume fraction $\phi$ was then estimated using the relation

$$\phi = \frac{\mu_{CNC}/\rho_{CNC}}{\mu_{CNC}/\rho_{CNC} + \mu_w/\rho_w}$$

where $\mu_w = 1 - \mu_{CNC}$ is the mass fraction of water and $\rho_{CNC}$ and $\rho_w$ are the mass densities of CNC and water (taken to be 1600 and 1000 kg $m^{-3}$ respectively).

### Ultrasonication

The nanocrystal size was controlled by high-intensity sonication (Fisherbrand Ultrasonic disintegrator, 20 kHz, tip diameter 12.7 mm operating with a 2 second : 1 second ON:OFF cycle). Unless otherwise stated, samples were sonicated at 2 wt% CNC with a sample volume of 20 mL in a truncated centrifuge tube (Corning Falcon 50 mL), with the sonicator tip immersed to a depth of one-third of the sample volume. Samples were immersed in an ice bath throughout sonication to prevent sample heating, as de-sulphation of the CNCs surface charges has been observed in acidic suspensions at high temperatures[48]. For longer sonication treatments, the ice bath was replenished every five minutes.

### Transmission electron microscopy (TEM)

TEM samples were prepared by casting a droplet of 0.001 wt% CNC suspension in pH 3 sulphuric acid solution onto carbon-coated copper grid prepared by glow discharge and staining with uranyl acetate solution. TEM images were captured using a Talos F200X G2 microscope (FEI) operating at 200 kV and a CCD camera. The CNC outlines were traced in Fiji (ImageJ) software and analysed using the Shape Filter plug-in. As CNCs exhibit considerable size polydispersity, at least 250 particles were measured for each sample to ensure reliable statistics. Discussion of the tracing protocol is given in section S1.10 of the Supplementary Information.

**Preparation of photonic CNC films**

To standardise the CNC suspensions after dialysis and ensure optimal visual appearance of the final films, each suspension was diluted to 1.6 wt% CNC in 1.92 mM NaCl solution. Films were cast from 2.5 mL of each suspension into a 35 mm diameter polystyrene Petri dish and left to dry under ambient conditions (20 °C, 40-50% relative humidity), which typically took 36 hours.

**Polarised optical microscopy**

Optical microscopy was performed on a Zeiss Axio microscope, with a halogen lamp (Zeiss HAL100) as light source using Koehler illumination. Images were captured in bright field reflection mode using a 10x objective (Nikon T Plan SLWD, NA 0.2) and recorded using CMOS camera (UI-3580LE-C-HQ, IDS). The white balance of the images was calibrated using a white Lambertian diffuser.

**Chiral nematic phase and pitch measurement**

CNC suspensions at a range of concentrations were prepared by first concentrating the stock using a rotary evaporator (40 °C, pressure < 40 mbar), then diluting to the desired final concentration using deionised water. Chiral nematic phase separation was observed in CNC suspensions at a range of concentrations. Glass capillaries were filled with the suspensions and sealed with UV epoxy (Norland Optical Adhesive 81), then left to equilibrate for at least two weeks. The anisotropic phase fraction was determined from the proportion of the suspension that appeared bright when viewed between crossed polarisers. The chiral nematic pitch of the anisotropic phase was measured as twice the periodicity of the fingerprint pattern observed in optical microscope images, which were collected using transmission illumination with crossed polarisers on a Zeiss Axioscope microscope equipped with a 20x objective (Nikon T Plan SLWD, NA 0.3).

**Acknowledgments**

**Funding**

This work was supported by the following funds: BBSRC [BB/V00364X/1]; EPSRC [EP/K503757/1, EP/P030467/1, EP/N509620/1, EP/L015978/1]; Philip Leverhulme Prize [PLP-2019-271]; ERC Horizon 2020 Framework Programme [Marie Curie Individual Fellowship 893136-MFCPF, ERC SeSaME ERC-2014-STG H2020 639088, ERC-2017-POC 790518, ERC BiTe ERC-2020-CoS-101001637, ITN-H2020 Plamatsu 722842]; Emil Aaltonen Foundation; Lord Lewis Research Studentship in Chemistry.

**Author contributions**

TGP, RMP, BFP and SV conceived the project and designed the experiments. TGP and GTV performed TEM imaging. JAH performed cryoTEM imaging. TGP and AN performed optical microscopy. TGP performed all other experiments and all data analysis. TGP, RMP and BFP interpreted the data. TGP wrote the manuscript with contributions from all authors.

**Competing interests**

Authors declare that they have no competing interests.

**Data and materials availability**

Additional data relating to this publication is available from the University of Cambridge data repository.

**Supplementary Information** Supplementary Methods, Figs. S1 to S21, Tables S1 to S7

## References

1. Jiang, W. *et al.* Emergence of complexity in hierarchically organized chiral particles. *Science* **368**, 642–648 (2020).
2. Zion, M. Y. B. *et al.* Self-assembled three-dimensional chiral colloidal architecture. *Science* **358**, 633–636 (2017).
3. Nyström, G., Arcari, M. & Mezzenga, R. Confinement-induced liquid crystalline transitions in amyloid fibril cholesteric tactoids. *Nature Nanotechnology* **13**, 330 (2018).
4. Siavashpouri, M. *et al.* Molecular engineering of chiral colloidal liquid crystals using DNA origami. *Nature Materials* **16**, 849–856 (2017).
5. Tortora, M. M. C., Mishra, G., Prešern, D. & Doye, J. P. K. Chiral shape fluctuations and the origin of chirality in cholesteric phases of DNA origamis. *Science Advances* **6**, eaaw8331 (2020).
6. Straley, J. P. Theory of piezoelectricity in nematic liquid crystals, and of the cholesteric ordering. *Phys. Rev. A* **14**, 1835–1841 (1976).

7. Grelet, E. & Fraden, S. What Is the Origin of Chirality in the Cholesteric Phase of Virus Suspensions? *Phys. Rev. Lett.* **90**, 198302 (2003).

8. Revol, J.-F., Bradford, H., Giasson, J., Marchessault, R. H. & Gray, D. G. Helicoidal self-ordering of cellulose microfibrils in aqueous suspension. *International Journal of Biological Macromolecules* **14**, 170–172 (1992).

9. Parker, R. M. *et al.* The Self-Assembly of Cellulose Nanocrystals: Hierarchical Design of Visual Appearance. *Adv. Mater.* **30**, 1704477 (2017).

10. Schütz, C. *et al.* From Equilibrium Liquid Crystal Formation and Kinetic Arrest to Photonic Bandgap Films Using Suspensions of Cellulose Nanocrystals. *Crystals* **10**, 199 (2020).

11. Shopsowitz, K. E., Qi, H., Hamad, W. Y. & MacLachlan, M. J. Free-standing mesoporous silica films with tunable chiral nematic structures. *Nature* **468**, 422–425 (2010).

12. Revol, J.-F., Godbout, J. & Gray, D. G. Solid self-assembled films of cellulose with chiral nematic order and optically variable properties. *J. Pulp Paper Sci.* **24**, 146–149 (1998).

13. Droguet, B. E. *et al.* Large-scale fabrication of structurally coloured cellulose nanocrystal films and effect pigments. *Nat. Mater.* 1–7 (2021) doi:10.1038/s41563-021-01135-8.

14. Hanley, S. J., Revol, J.-F., Godbout, L. & Gray, D. G. Atomic force microscopy and transmission electron microscopy of cellulose from Micrasterias denticulata; evidence for a chiral helical microfibril twist. *Cellulose* **4**, 209–220 (1997).

15. Majoinen, J. *et al.* Supracolloidal Multivalent Interactions and Wrapping of Dendronized Glycopolymers on Native Cellulose Nanocrystals. *J. Am. Chem. Soc.* **136**, 866–869 (2014).

16. Ogawa, Y. Electron microdiffraction reveals the nanoscale twist geometry of cellulose nanocrystals. *Nanoscale* **11**, 21767–21774 (2019).

17. Paavilainen, S., Róg, T. & Vattulainen, I. Analysis of Twisting of Cellulose Nanofibrils in Atomistic Molecular Dynamics Simulations. *J. Phys. Chem. B* **115**, 3747–3755 (2011).

18. Zhao, Z. *et al.* Cellulose Microfibril Twist, Mechanics, and Implication for Cellulose Biosynthesis. *J. Phys. Chem. A* **117**, 2580–2589 (2013).

19. Dumitrică, T. Intrinsic twist in Iβ cellulose microfibrils by tight-binding objective boundary calculations. *Carbohydrate Polymers* **230**, 115624 (2020).

20. Khandelwal, M. & Windle, A. Origin of chiral interactions in cellulose supra-molecular microfibrils. *Carbohydrate Polymers* **106**, 128–131 (2014).

21. Habibi, Y., Lucia, L. A. & Rojas, O. J. Cellulose Nanocrystals: Chemistry, Self-Assembly, and Applications. *Chem. Rev.* **110**, 3479–3500 (2010).

22. de Gennes, P. G. & Prost, J. *The Physics of Liquid Crystals*. (Clarendon Press, 1995).

23. Beck, S., Bouchard, J. & Berry, R. Controlling the Reflection Wavelength of Iridescent Solid Films of Nanocrystalline Cellulose. *Biomacromolecules* **12**, 167–172 (2011).

24. Heux, L., Chauve, G. & Bonini, C. Nonflocculating and chiral-nematic self-ordering of cellulose microcrystals suspensions in nonpolar solvents. *Langmuir* **16**, 8210–8212 (2000).

25. Elazzouzi-Hafraoui, S. *et al.* The Shape and Size Distribution of Crystalline Nanoparticles Prepared by Acid Hydrolysis of Native Cellulose. *Biomacromolecules* (2007) doi:10.1021/bm700769p.

26. Neto, W. P. F. *et al.* Comprehensive morphological and structural investigation of cellulose I and II nanocrystals prepared by sulphuric acid hydrolysis. *RSC Adv.* **6**, 76017–76027 (2016).

27. Araki, J. & Kuga, S. Effect of Trace Electrolyte on Liquid Crystal Type of Cellulose Microcrystals. *Langmuir* **17**, 4493–4496 (2001).

28. Dussi, S. & Dijkstra, M. Entropy-driven formation of chiral nematic phases by computer simulations. *Nature Communications* **7**, 11175 (2016).

29. Terech, P., Chazeau, L. & Cavaille, J. Y. A Small-Angle Scattering Study of Cellulose Whiskers in Aqueous Suspensions. *Macromolecules* **32**, 1872–1875 (1999).

30. Usov, I. *et al.* Understanding nanocellulose chirality and structure–properties relationship at the single fibril level. *Nat Commun* **6**, 7564 (2015).

31. Uhlig, M. *et al.* Two-Dimensional Aggregation and Semidilute Ordering in Cellulose Nanocrystals. *Langmuir* **32**, 442–450 (2016).

32. Onsager, L. The Effects of Shape on the Interaction of Colloidal Particles. *Annals of the New York Academy of Sciences* **51**, 627–659 (1949).

33. Xu, Y., Atrens, A. & Stokes, J. R. A review of nanocrystalline cellulose suspensions: Rheology, liquid crystal ordering and colloidal phase behaviour. *Advances in Colloid and Interface Science* **275**, 102076 (2020).

34. Honorato-Rios, C. & Lagerwall, J. P. F. Interrogating helical nanorod self-assembly with fractionated cellulose nanocrystal suspensions. *Communications Materials* **1**, 1–11 (2020).

35. Honorato-Rios, C. *et al.* Fractionation of cellulose nanocrystals: enhancing liquid crystal ordering without promoting gelation. *NPG Asia Materials* **10**, 455–465 (2018).

36. Nyström, G., Arcari, M., Adamcik, J., Usov, I. & Mezzenga, R. Nanocellulose Fragmentation Mechanisms and Inversion of Chirality from the Single Particle to the Cholesteric Phase. *ACS Nano* **12**, 5141–5148 (2018).

37. Chiappini, M., Dussi, S., Frka-Petesic, B., Vignolini, S. & Dijkstra, M. Modeling the cholesteric pitch of apolar cellulose nanocrystal suspensions using a chiral hard-bundle model. *J. Chem. Phys.* **156**, 014904 (2022).

38. Majoinen, J. *Synthetic and Supracolloidal Concepts for Cellulose Nanocrystals*. (Aalto University, 2016).

39. Nemati, A. *et al.* Effects of shape and solute-solvent compatibility on the efficacy of chirality transfer: Nanoshapes in nematics. *Science Advances* (2022) doi:10.1126/sciadv.abl4385.

40. Gonçalves, D. P. N. & Hegmann, T. Chirality Transfer from an Innately Chiral Nanocrystal Core to a Nematic Liquid Crystal: Surface-Modified Cellulose Nanocrystals. *Angewandte Chemie International Edition* **60**, 17344–17349 (2021).

41. Dong, X. M., Kimura, T., Revol, J.-F. & Gray, D. G. Effects of Ionic Strength on the Isotropic-Chiral Nematic Phase Transition of Suspensions of Cellulose Crystallites. *Langmuir* **12**, 2076–2082 (1996).

42. Narkevicius, A. *et al.* Controlling the Self-Assembly Behavior of Aqueous Chitin Nanocrystal Suspensions. *Biomacromolecules* **20**, 2830–2838 (2019).

43. Bai, L. *et al.* Chirality from Cryo-Electron Tomograms of Nanocrystals Obtained by Lateral Disassembly and Surface Etching of Never-Dried Chitin. *ACS Nano* **14**, 6921–6930 (2020).

44. Reid, M. S., Villalobos, M. & Cranston, E. D. Benchmarking Cellulose Nanocrystals: From the Laboratory to Industrial Production. *Langmuir* (2016) doi:10.1021/acs.langmuir.6b03765.

45. Delepierre, G., Vanderfleet, O. M., Niinivaara, E., Zakani, B. & Cranston, E. D. Benchmarking Cellulose Nanocrystals Part II: New Industrially Produced Materials. *Langmuir* **37**, 8393–8409 (2021).

46. Calvaresi, M. The route towards nanoparticle shape metrology. *Nat. Nanotechnol.* **15**, 512–513 (2020).

47. Boselli, L. *et al.* Classification and biological identity of complex nano shapes. *Communications Materials* **1**, 1–12 (2020).

48. Beck, S. & Bouchard, J. Auto-catalyzed acidic desulfation of cellulose nanocrystals. *Nordic Pulp and Paper Research Journal* **29**, 006–014 (2014).

49. *Cotton plant photo credit: Seven Seven, unsplash.com/photos/yvmg-tuc7es . Reproduced under Unsplash license.*

## Figures

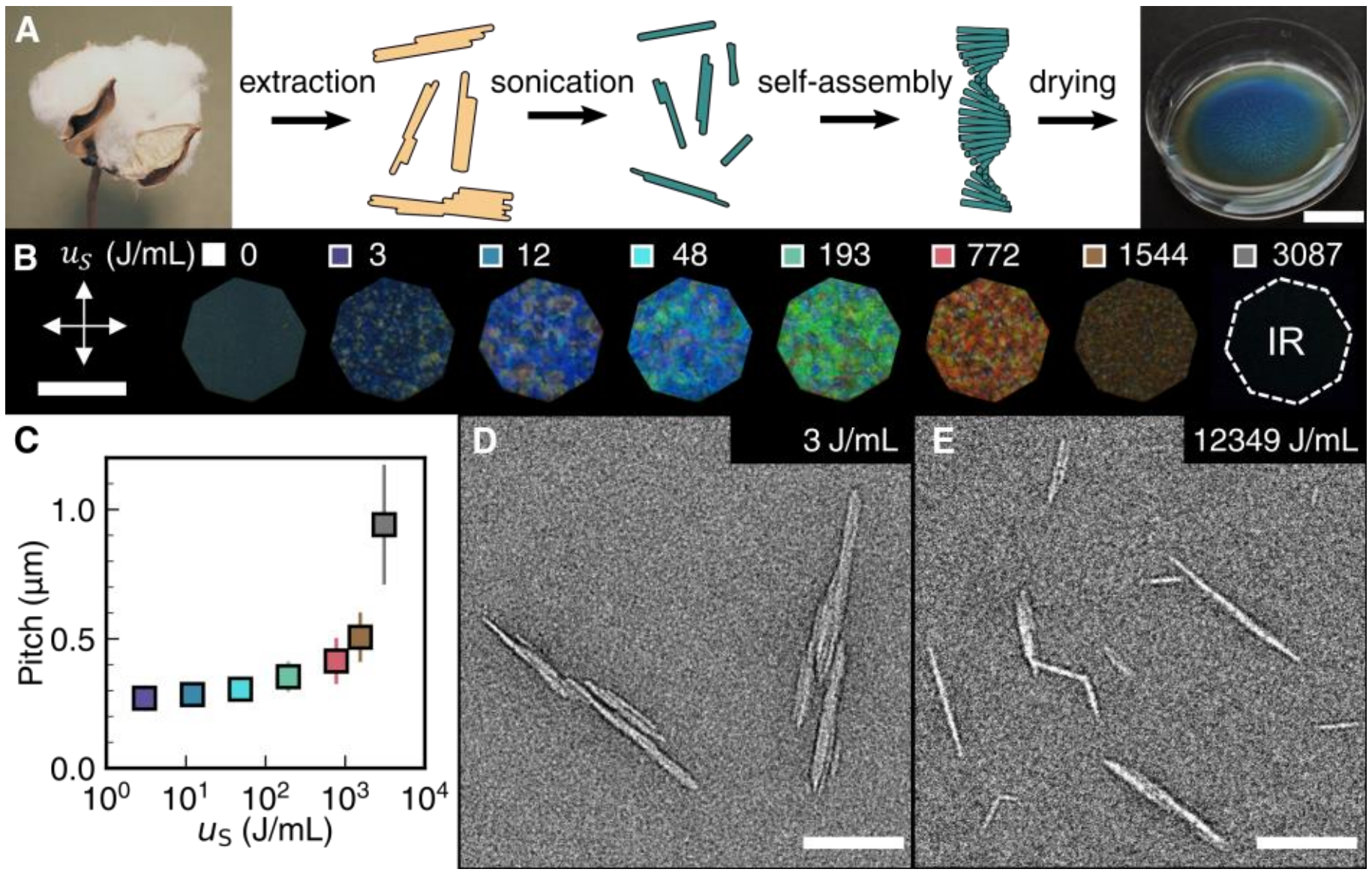

**Fig. 1. Tuning the colour of photonic CNC films with sonication** (**A**) Formation of structurally coloured films via chiral self-assembly of CNCs extracted from cotton. Cotton image source: [49]. (**B**) Photonic CNC films of increasing sonication dose $u_S$ viewed under crossed polarisers. Square panels next to each dose value indicate the colour code used for other plots. IR = infrared. Scale bar 100 µm. (**C**) Pitch increase with sonication dose, as inferred from cross-sectional scanning electron microscopy. Error bars indicate standard deviation of 30 measurements. (**D-E**) Examples of TEM images of CNCs, showing their composite morphology. Scale bar 100 nm.

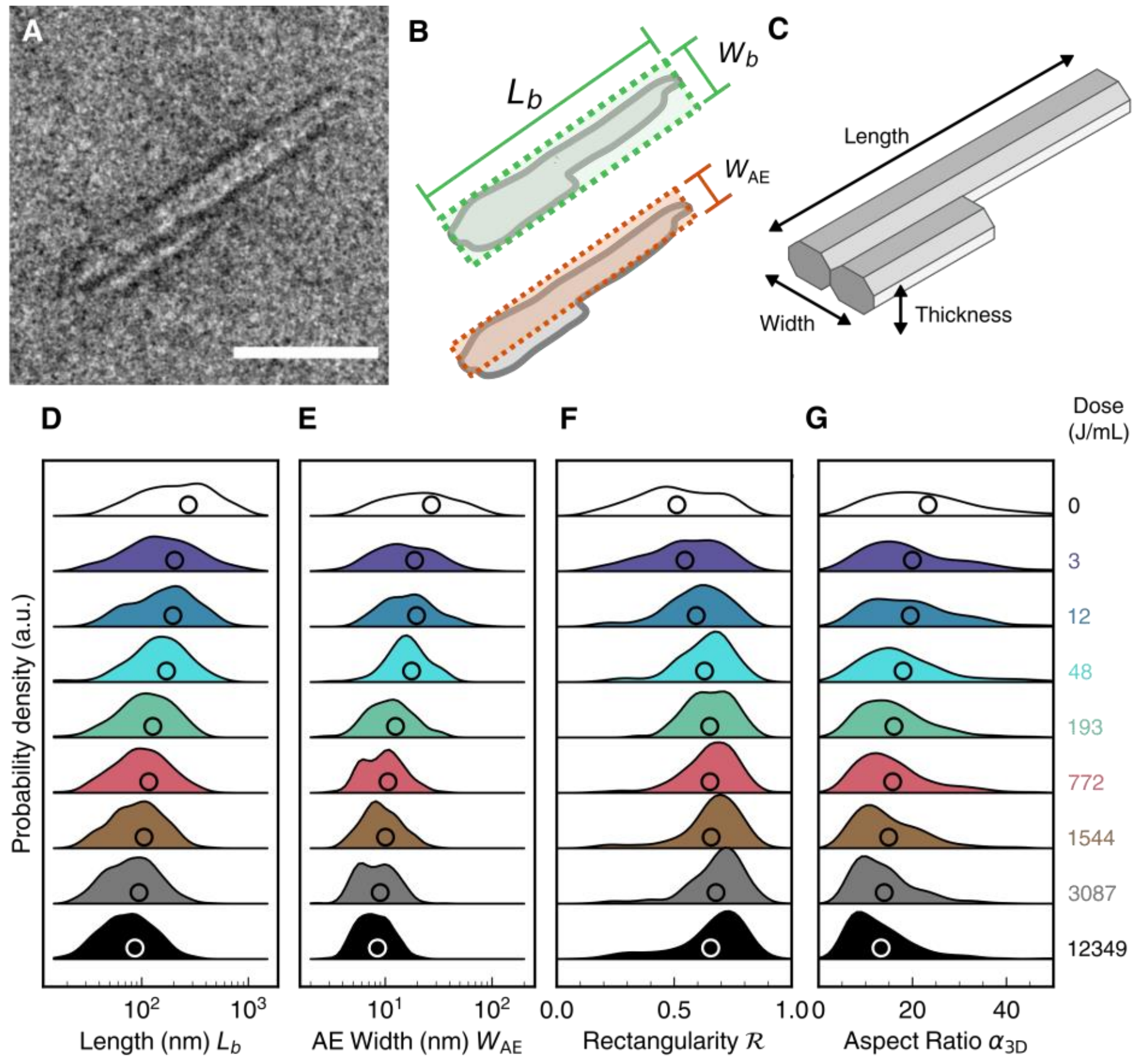

**Fig. 2. Revealing the 3D morphological changes due to sonication by nanoparticle shape metrology** (**A**) Example of a CNC particle in a TEM image. Scale bar 50 nm. (**B**) Definition of the box length, box width, and area-equivalent (AE) width. (**C**) Schematic of the 3D morphology of the particle in (A), with the length, width and thickness dimensions indicated (**D-G**) Distributions in particle size and shape properties for box length (D), AE width (E), rectangularity (F) and 3D aspect ratio (G), plotted in vertical stacks to facilitate comparison of statistical features between samples. Open circles indicate mean values.

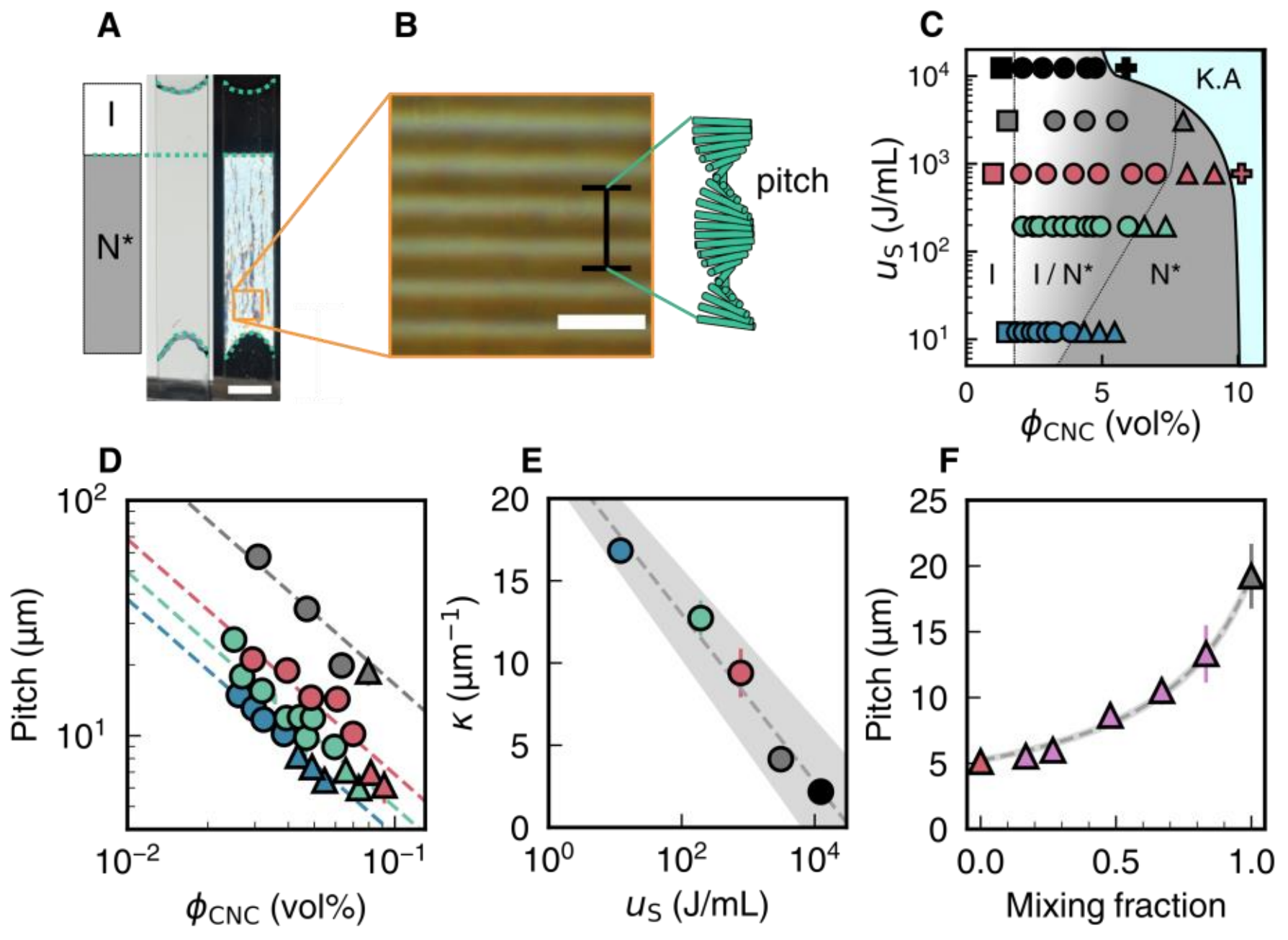

**Fig. 3 Quantitative analysis of chiral self-assembly behaviour** (**A**) Macroscopic phase separation of a biphasic CNC suspension in a sealed rectangular glass capillary. Vertical separation into isotropic (I) and chiral nematic (N*) phases is evident when viewed under crossed polarisers (right photograph). Scale bar 5 mm. (**B**) Polarised optical microscopy of the N* phase from (A). The periodicity of the fingerprint pattern corresponds to a 180˚ rotation of the chiral nematic order and is equal to half the pitch. Scale bar 10 µm. (**C**) Phase diagram versus CNC concentration and sonication dose, with the isotropic (I, squares), biphasic (I/N*, circles), chiral nematic (N*, triangles) and kinetically arrested (K.A., crosses) phases indicated. (**D**) Variation in pitch with CNC concentration for biphasic and chiral nematic suspensions (symbols match the phase diagram (C)). Dotted lines indicate linear fitting. (**E**) Decrease in chiral strength κ with sonication dose. Error bars indicate standard deviation of 30 measurements. Grey line indicates empirical logarithmic relation. Grey shaded region indicates uncertainty in fitting. (**F**) Pitch of mixtures of 'low dose' (772 J/mL, red) and 'high dose' (3087 J/mL, brown) chiral nematic suspensions at equal CNC volume fraction. Error bars indicate standard deviation of 30 measurements. Grey line indicates fitting to chiral dopant model. Grey shaded region indicates uncertainty in fitting.

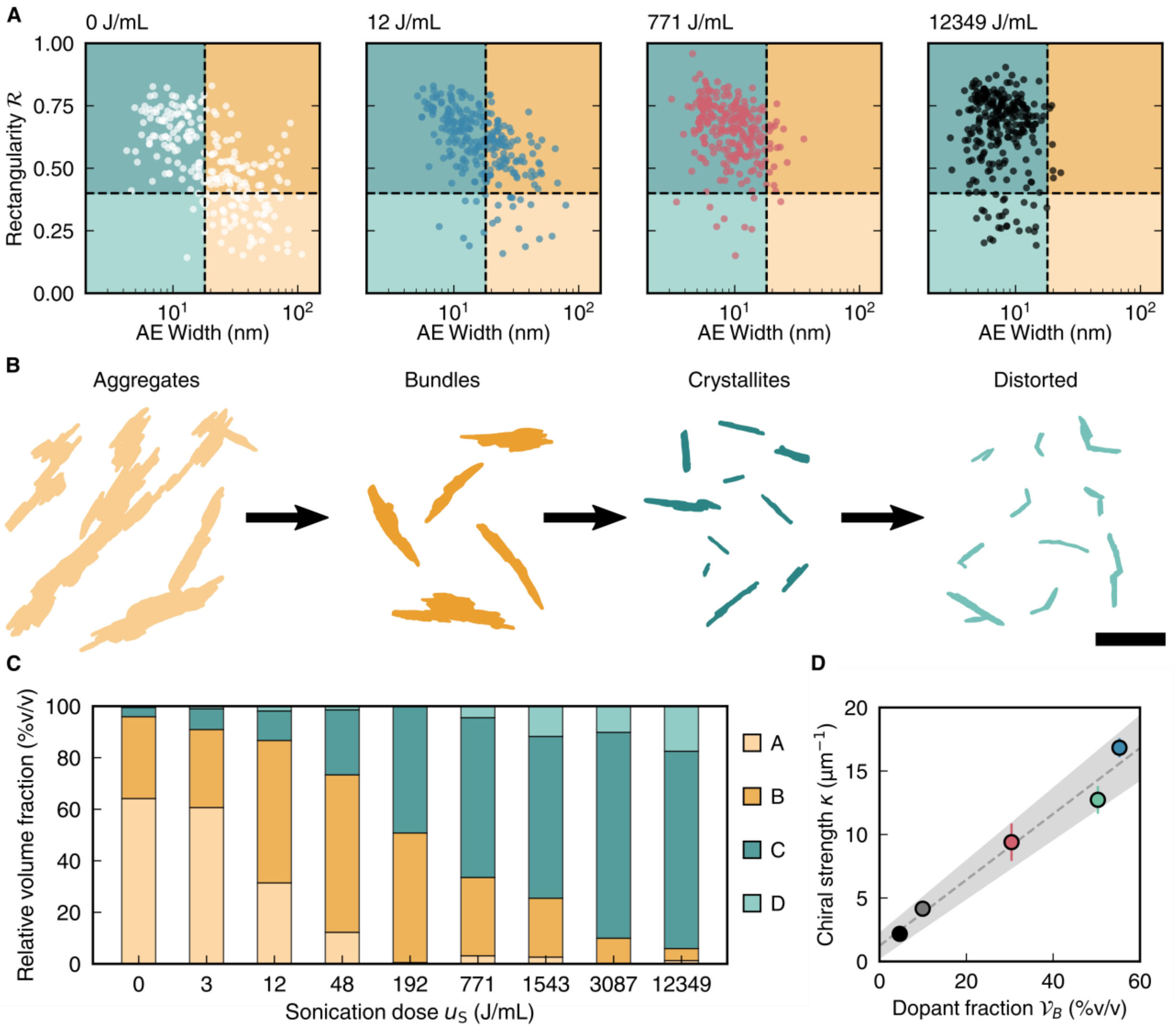

**Fig. 4 Relating the CNC mesophase chirality to the abundance of crystallite bundles** (**A**) Classification of CNC particles into four sub-populations based on AE width and rectangularity. Scatter plots of AE width versus rectangularity are shown for four sonication doses, illustrating the evolution of the system under sonication. (**B**) Representative examples of each particle sub-population. Evolution under sonication indicated by black arrows. Scale bar 200 nm. (**C**) Relative volume fraction of each particle sub-population versus sonication dose. (**D**) Modulation of chiral strength with relative dopant volume fraction. Error bars indicate standard deviation of 30 measurements. Grey dotted line indicates linear correlation. Grey shaded region indicates uncertainty in fitting.

# Supplementary Information for Chiral Self-Assembly of Cellulose Nanocrystals is Driven by Crystallite Bundles

Thomas G. Parton, Richard M. Parker, Gea T. van de Kerkhof,
Aurimas Narkevicius, Johannes S. Haataja, Bruno Frka-Petesic, Silvia Vignolini*
Yusuf Hamied Department of Chemistry, University of Cambridge

*Corresponding author sv319@cam.ac.uk

# Contents

# S1 Supplementary Methods

## S1.1 Dynamic light scattering (DLS)

DLS samples were prepared with CNC (0.1 wt%) suspended in 1 mM NaCl solution, and passed through a 0.8 µm cellulose acetate syringe filter before pouring into disposable plastic cuvettes. DLS measurements were performed with a commercial setup (Malvern Zetasizer Nano ZS) using 633 nm illumination and collecting back-scattered (173°) light. Samples were measured at 25 °C after a delay of 300 s to allow for thermal equilibration before data collection, which was performed in three runs of at least ten measurements per run.

## S1.2 Electrolytic conductivity and pH measurement

The conductivity and pH of CNC suspensions were determined using a platinum 2-pole conductivity probe (InLab 752-6MM, Mettler Toledo) or pH probe (InLab Micro Pro-ISM, Mettler Toledo) respectively. All measurements were performed at room temperature.

## S1.3 Optical spectroscopy of CNC photonic films

Left-circular polarised (LCP) reflectance spectra of CNC photonic films were obtained using the microscope setup used to collect polarised optical microscopy images (see Methods and Figure 1 of the main article). The light reflected from the sample passed through an LCP analyser, composed of an achromatic quarter-wave plate and linear polariser (Thorlabs WP25M-UB), before being transmitted to a UV-vis spectrometer (AvaSpec-HS2048, Avantes) *via* an optical fibre (600 µm core diameter, FC-UV600-2-SR, Avantes). Reflection spectra were normalised to a silver mirror (Thorlabs, PF10-03-P01).

## S1.4 Scanning electron microscopy (SEM) of CNC photonic films

Fully dry CNC films were pulled apart to expose their cross-sections. The films were then mounted onto steel stubs using conductive carbon tape. Samples were coated with platinum to a nominal thickness of 10 nm using a sputter coater (Quorum Q150T ES) to ensure good conduction of electrons. Micrographs were taken using a TESCAN MIRA3 FEG-SEM system using an acceleration voltage of 5 kV and a working distance of 3-6 mm.

## S1.5 Conductometric titration of CNC suspensions

In a typical titration procedure, 2.0 g of 2.0 wt% CNC suspension was added to 200 mL of 0.5 mM NaCl solution. The acidic CNC suspensions (with $H^+$ counter-ions on the CNC sulfate half-ester groups) underwent extensive dialysis against deionised water prior to measurement. An automatic titrator (Metrohm 856) was used to inject NaOH solution (10 mM) in 5 µL increments, while continuously recording the suspension conductivity. The surface charge per CNC dry mass (mmol/kg) was determined from the first equivalence point of the titration curve, obtained by a manual piecewise linear fitting.

## S1.6 Zeta potential of CNC suspensions

The CNC samples used for zeta potential measurements were identical to those used for DLS measurements (section S1.1). The zeta potential was estimated from the measured electrophoretic mobility using the Smoluchowski limit of the Henry equation ($\kappa R \gg 1$, $F(\kappa R) = 1.5$), which is the conventional choice for CNC suspensions (1). Measurements were performed in three batches of at least 50 runs each.

## S1.7 UV-vis transmission spectroscopy of CNC suspensions

UV-vis transmission spectra were obtained using a commercial spectrophotometer (Cary 4000). CNC samples were prepared at 0.1 wt% and measured in a quartz cuvette (Hellma 100-10-40) with 10 mm path length.

## S1.8 Cryogenic transmission electron microscopy (cryoTEM) of CNC suspensions

Cryogenic transmission electron microscopy (cryoTEM) imaging was carried out using a JEM 3200FSC field emission microscope (JEOL) operated at 300 kV in bright field mode with an Omega-type zero-loss energy filter. The images were acquired with a Ultrascan 4000 CCD camera (Gatan) and processed with Gatan Digital Micrograph software (version 1.83.842).

Vitrified samples were prepared using EM GP2 Automatic Plunge Freezer by placing a 4-5 µL droplet of sample solution onto plasma-cleaned 300-mesh lacey carbon copper grids in a 90% humidity atmosphere, then blotted with filter paper for 0.5 - 1.5 seconds, followed by immediate immersion into an ethane/propane mixture at −170 °C. The samples were then cryo-transferred to the microscope, where their temperature was maintained at −187 °C.

## S1.9 Atomic force microscopy (AFM) of CNC suspensions

Atomic force microscopy (AFM) images of the cellulose nanocrystals were acquired at ambient conditions using a scanning probe microscope (Agilent 5500 SPM) in tapping mode with an AFM probe (OTESPA-R3). A square of mica (2 $cm^2$) was freshly cleaved to obtain a mirror smooth surface. A droplet (100 µL, 0.1 wt%) of poly-L-lysine (P8920, Sigma, $M_W$ = 150-300 kDa) was deposited for 1 minute, after which it was rinsed off with deionised water and dried under nitrogen gas flow. Then, the CNC sample (150 µL, 0.001 wt%, pH = 3) was deposited and incubated for 3 minutes after which it was washed off with deionised water and dried under nitrogen gas flow. Finally, the sample was dried in the oven for 30 minutes at 50 °C and then kept at ambient conditions in a closed dish before the measurements. Scans were typically performed over a 4x4 $\mu m^2$ area with 2048 points per line at 0.6 Hz to acquire the final images with ~2 nm resolution.

CNC particle height statistics were extracted from AFM images using Gwyddion software (2). The images were processed in six steps: (1) removing the background height variation using a fifth-order polynomial fitting (2) aligning rows by median values (3) healing scars (4) flattening the base (5) setting the zero offset to the median image height and finally (6) Gaussian filtering with 2 pixel resolution.

To estimate the AFM tip diameter, AFM images were also obtained for spherical gold nanoparticles with diameter $D \approx 15$ nm. The tip diameter was estimated by the difference between the max height of the particle and the apparent lateral diameter (Figure S11C). Measurement on 38 particles gave $D_{\mathrm{tip}}$ = 14.2 ± 1.1 nm (Figure S1)

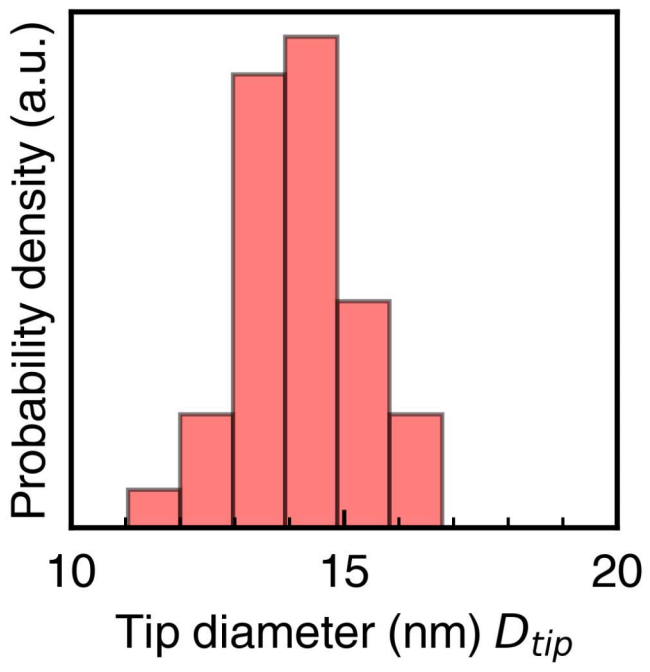

**Figure S1:** Histogram of estimated tip diameter $D_{\mathrm{tip}} = D_{\mathrm{eff}} - D$ for spherical gold nanoparticles.

### S1.10 Transmission electron microscopy (TEM) shape analysis

As previous authors have noted, there is great variability in CNC size data obtained from TEM images, even for two users manually measuring identical experimental data (3). The tracing method used in this work is therefore described in detail below to aid comparison with other works.

CNC particles as observed in TEM images were manually traced to extract size and shape properties. Tracing was performed on a touch-screen device using a stylus and the "Freehand Selections" tool in Fiji/imageJ. In negatively-stained TEM images, CNCs generally appear as bright spindle-shaped objects surrounded by a dark "halo" or outline created by the accumulation of staining agent around the particle. The CNC particle was taken to be the entire bright area within this dark outline. The particle areas were filled white on the original image and then selected by thresholding the image.

We found that suspending the CNCs in an aqueous pH 3 solution of sulphuric acid dramatically reduced the occurrence of artefacts due to particle aggregation on the dried TEM grid, in agreement with previous work (4). Therefore, any single continuous bright region, regardless of shape, was assumed to be a single CNC particle. In particular, even if the shape could be interpreted as two elongated particles overlapping, the shape was still assumed to be a single particle.

The traced shapes were analysed using the ImageJ Shape Filter plugin, filtering for all particles larger than 70 $\mathrm{nm}^2$ to eliminate any artefacts due to pixel noise. The data were exported in .csv format and further processed using a custom Python script.

## S2 Calibration of sonication dose by calorimetry

The impact of ultrasonication of CNC suspensions depends on the properties of the suspension (volume and CNC concentration), the settings on the tip sonicator (tip amplitude) and the duration of the treatment (Figure S2A). Although sonication is frequently applied to CNC suspensions and the equipment settings are often provided in the literature, it can be difficult to make quantitative comparisons between results. To explore the consistency of sonication across different experimental conditions, the hydrodynamic diameter of CNC particles after various sonication treatments was measured using DLS (section S1.1). The difference treatments are summarised in Figure S2B).

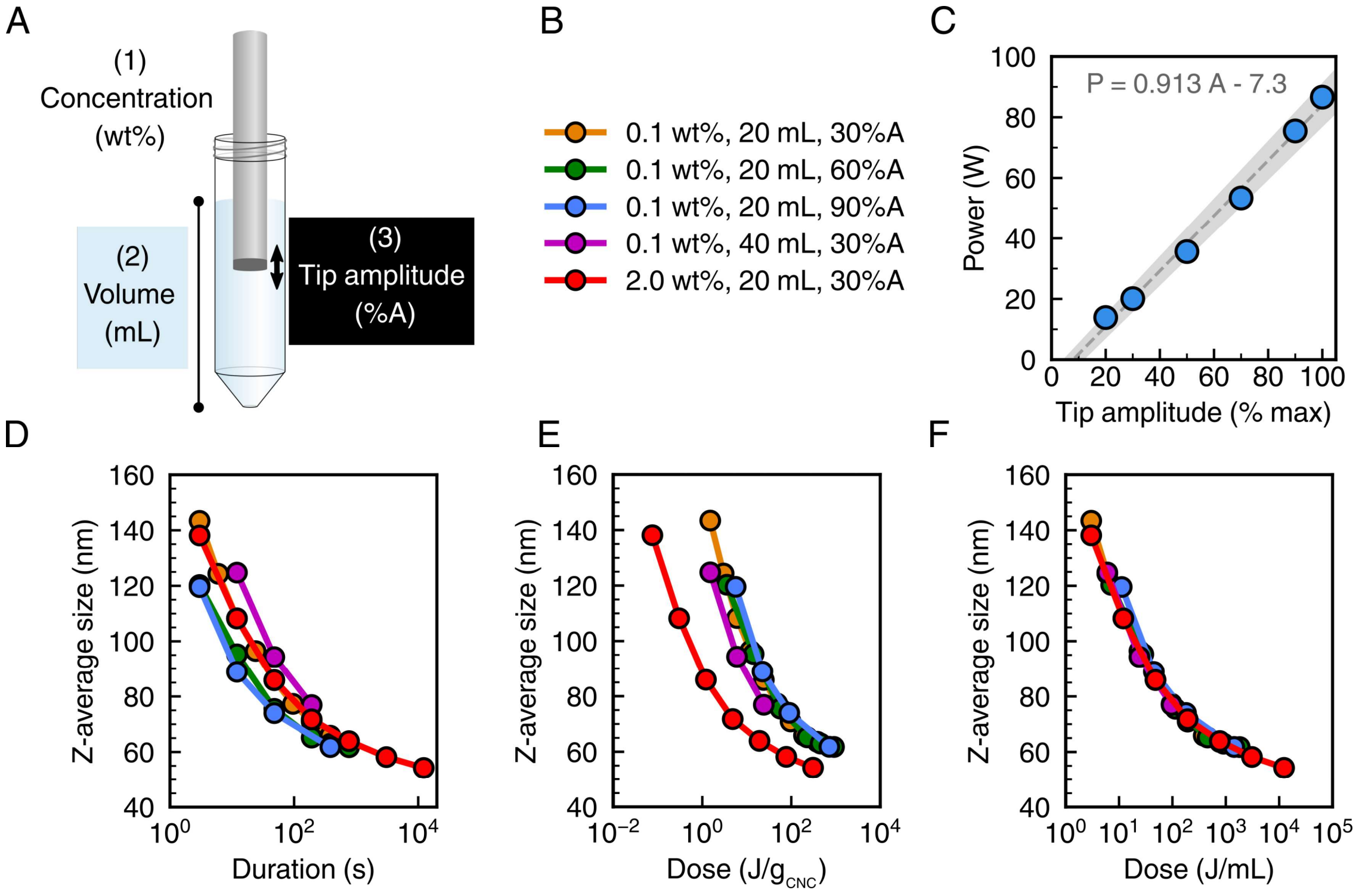

**Figure S2:** (**A**) Schematic of experimental setup for tip sonication, with key experimental parameters indicated. (**B**) List of experimental conditions used for calibration experiments. Colour scheme matches that used in (**D**-**F**). (**C**) Caloric power delivered to suspension versus tip amplitude, with linear fitting. (**D**-**F**) DLS z-average size versus various dose parameters: (D) duration of treatment, (E) energy per CNC dry mass $w_S$, and (F) energy per suspension volume $u_S$.

To quantify the impact of sonication in a device-independent manner, it is necessary to express the sonication dose as the total energy delivered to the system by the sonicator tip. As sonication increases the temperature of the suspension, the energy delivered to the suspension can be estimated by calorimetry. Calorimetry was performed using a custom-made bomb calorimeter, consisting of a cylindrical chamber of expanded polystyrene foam surrounding a centrifuge tube to thermally insulate the sample. The temperature was measured using a thermocouple (Thorlabs TSP01) within the chamber. A magnetic stirrer bar was used to maintain mixing of the suspension.

First, the power $P$ delivered to a sample of 40 g deionised water was determined for a range of values for the amplitude of the sonicator tip (expressed as a percentage of the maximum amplitude, $\%A$), as shown in Figure S2C. Using this calibration curve for $P(\%A)$, the sonication energy $E$ delivered to the sample is given by

$$E(\%A) = P(\%A)t \tag{1}$$

where $t$ is the total duration of the sonication treatment (considering only the duration of the ON portion of the ON:OFF sonication cycles). This expression assumes that the CNC suspensions have the same specific heat capacity as water, which is a valid assumption at low CNC concentration.

The decrease in hydrodynamic diameter (DLS z-average size) with sonication is shown in Figure S2D-F, where the sonication dose is expressed in terms of sonication duration $t$ (Figure S2D), energy per CNC dry mass ($w_S = E/m_{\mathrm{CNC}}$, Figure S2E) and energy per suspension volume ($u_S = E/V$, Figure S2F).

It is clear from Figure S2D-F that the hydrodynamic diameter results are only consistent across all experimental conditions when the sonication dose is expressed using $u_S$. Note that energy per CNC dry mass $w_S$, a dose unit often used in the CNC literature, gives consistent results when comparing conditions at fixed CNC concentration, but not when comparing doses at different concentrations (Figure S2E).

# S3 Discussion of other explanations of pitch increase induced by sonication

## S3.1 Release of trapped ions

The breakdown of CNCs is associated with the release of trapped ions, increasing the electrolytic conductivity of the suspension. The conductivity and pH of sonicated CNC suspensions were measured after sonication but before dialysis (section S1.2), producing the results shown in table S1.

| Sonication dose (J/mL) | Conductivity ($\mu$S cm$^{-1}$) | pH |
|---:|---:|---:|
| 0 | 362.1 ± 0.2 | 2.76 |
| 12 | 428.0 ± 0.4 | 2.71 |
| 48 | 500.9 ± 1.0 | 2.67 |
| 193 | 802.0 ± 0.9 | 2.55 |
| 772 | 1535.9 ± 1.8 | 2.34 |

**Table S1:** Table of electrolytic conductivity and pH for sonicated CNC suspensions before dialysis.

The decrease in pH with sonication indicates that $H^+$ ions are released, and the associated contribution to the conductivity increase can be estimated from the pH shift. Explicitly, denoting the conductivity and pH of the never-sonicated suspension as $S_0$ and $\mathrm{pH}_0$ respectively, the expected conductivity contribution from $H^+$ ions is

$$\Delta S_{H^+} = \Lambda_{H^+}(10^{-\mathrm{pH}} - 10^{-\mathrm{pH}_0}) \tag{2}$$

where ($\Lambda_{H^+} \approx 0.035\,\mathrm{m^2\,S\,mol^{-1}}$) is the ionic conductivity of $H^+$ in aqueous solution at infinite dilution (5). The estimated $H^+$ contribution compared to the experimental conductivity change is shown in fig. S3. These results indicate that the release of $H^+$ ions is the dominant source of conductivity change. It is likely that the released $H^+$ ions originate from the acid hydrolysis used to produce the CNCs.

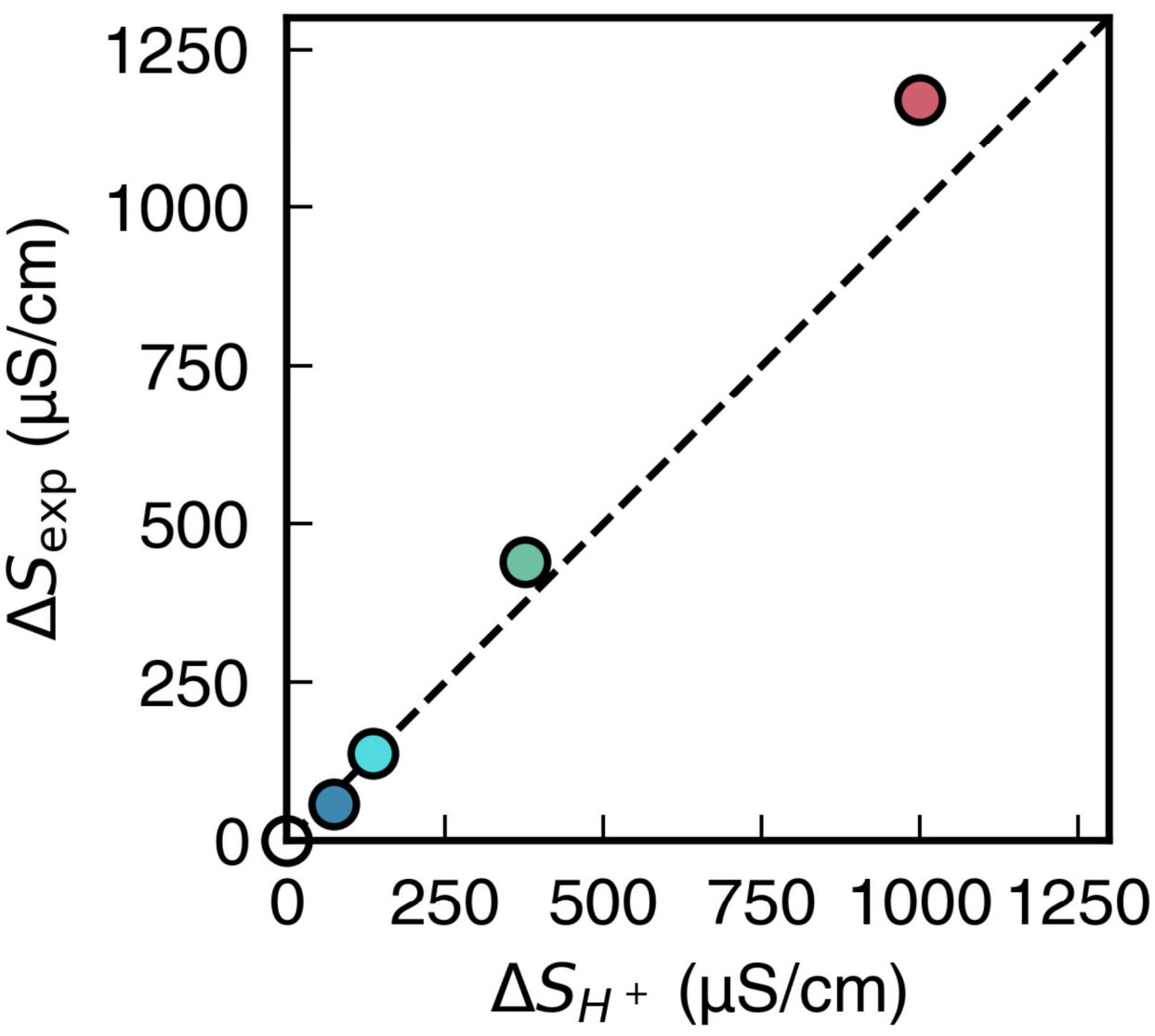

**Figure S3:** Comparison of estimated $H^+$ contribution to the conductivity change calculated using table S1 and eq. (2) versus experimentally-measured conductivity change. Dotted black line indicates equality.

It has been proposed that these released ions increase the effective particle volume and thereby cause a red-shift in the chiral nematic pitch by preventing short-range chiral interactions between CNCs (6). However, this argument seems to contradict the expected behaviour of charged colloids, where increasing ionic strength is usually found to reduce the thickness of the electric double layer (7). Furthermore, numerous studies have shown that increasing the ionic strength of liquid crystalline CNC suspensions results in a smaller pitch, leading to a blue-shifted colour in the photonic film (8). In this work, the suspensions were therefore extensively dialysed (> 1 week) after sonication, all in the same bath of deionised water, to rule out the influence of released ions. After dialysis, an apparent difference in conductivity was measured between samples and could not simply be attributed to variation in CNC concentration after dialysis (table S2)

| Sonication dose (J/mL) | Conductivity (µS $cm^{-1}$) | CNC concentration (wt%) |
|---|---|---|
| 0 | 192.5 ± 2.0 | 1.99 |
| 12 | 262.4 ± 0.8 | 2.03 |
| 48 | 254.8 ± 1.8 | 1.80 |
| 193 | 343.2 ± 7.2 | 1.88 |
| 772 | 380.3 ± 2.8 | 1.93 |

**Table S2:** Table of electrolytic conductivity and CNC concentration values for sonicated CNC suspensions after dialysis.

If dialysis has reached completion, the only ions remaining in the CNC suspension should be $H^+$ counterions. However, tight binding of these counterions to the CNC surface is expected to reduce their effective mobility, and therefore their contribution to conductivity and pH measurements. The strength of this binding is expected to vary with sonication as the total CNC surface area increases (see section S3.2 below). To explore this effect, we followed a procedure recently reported to give more accurate pH values of CNC suspensions by preparing samples at fixed CNC concentration (1.50 wt%) with an excess of $K^+$ ions (50 mM KCl) to dislodge the $H^+$ ions from the CNC surface (9).

All CNC suspensions with added salt were found to have the same pH value within experimental uncertainty (2.56±0.03), demonstrating that the overall number of $H^+$ counterions per CNC mass was identical. This result is in agreement with surface charge values determined by conductometric titration (section S3.2), and provides direct evidence for a sonication-induced decrease in the effective binding strength of counter-ions to the CNC surface. Note that this trend is consistent with the increase in specific surface area with sonication (fig. S10C).

## S3.2 Surface charge and colloidal stability

The surface charge on CNCs influences their colloidal stability and self-assembly by modifying the electrostatic interactions between particles (10). The effect of sonication on the CNC surface charge was therefore investigated as a possible source of the observed variation in the chiral nematic pitch.

The CNC surface charge (expressed as moles of counterions per kilogram of CNC dry mass) was determined by conductometric titration (Section S1.5) after dialysis. As shown in Table S3 the surface charge per CNC dry mass did not vary with sonication dose within the uncertainty of the fitting of the titration curve. The surface charge for these samples ($\approx$ 153 mmol $kg^{-1}$) corresponds to a CNC sulphur content of 0.49% (w/w), or a degree of substitution of 2.5% on glucose monomers.

Although the CNC specific surface charge (i.e. surface charge per mass) $S$ is readily accessible by experiment, a more relevant physical properties is charge per surface area $\sigma$, also known as the areal surface charge density, which is given by $\sigma = S/\mathrm{SSA}$, where SSA is specific surface charge (surface area per CNC mass). The SSA for a CNC particle of surface area $\Sigma$ and volume $V$ is given by $\mathrm{SSA} = \Sigma/(\rho_{\mathrm{CNC}} V)$, where $\rho_{\mathrm{CNC}}$ is the CNC mass density and assumed to be 1600 kg $m^{-3}$. For a distribution of particles, the mean surface charge density can be estimated by assuming the specific surface charge is identical for all particles, i.e.

$$\langle \sigma \rangle = \langle \frac{S}{\mathrm{SSA}} \rangle$$

as opposed to assuming the charge per surface area is identical (which would give $\langle \sigma \rangle' = S/\langle \mathrm{SSA} \rangle$). The CNC surface charge density value is therefore highly dependent on the method used to estimate SSA.

Estimates of the CNC charge per surface area obtained by three possible methods are shown in fig. S4. The "TEM outline" estimate is obtained by calculating the average SSA using the SSA of each particle based on the surface area and volume estimates used elsewhere in this work (see section S7 and fig. S10C). The "box" estimate is obtained in a similar way, but assumes that the CNCs are cuboidal, with length, width and thickness given by their box length $L_b$, box width $W_b$ and mean thickness $\langle T \rangle$ respectively (see section S7 for definitions of these properties). The box estimate consistently produces a higher value for $\langle \sigma \rangle$, due to higher estimates of the particle volume. Alternatively, as shown in previous works (e.g. (10)), the CNC surface charge density can be estimated by assuming the CNCs are identical cylinders with length and diameter given by the box length and thickness. Using this "mean cylinder" approach, substantially lower estimates for $\langle \sigma \rangle$ were obtained. These results illustrate the considerable uncertainty in estimating the surface charge density of CNCs. Nevertheless, all estimates predict a decrease in $\langle \sigma \rangle$ with sonication dose

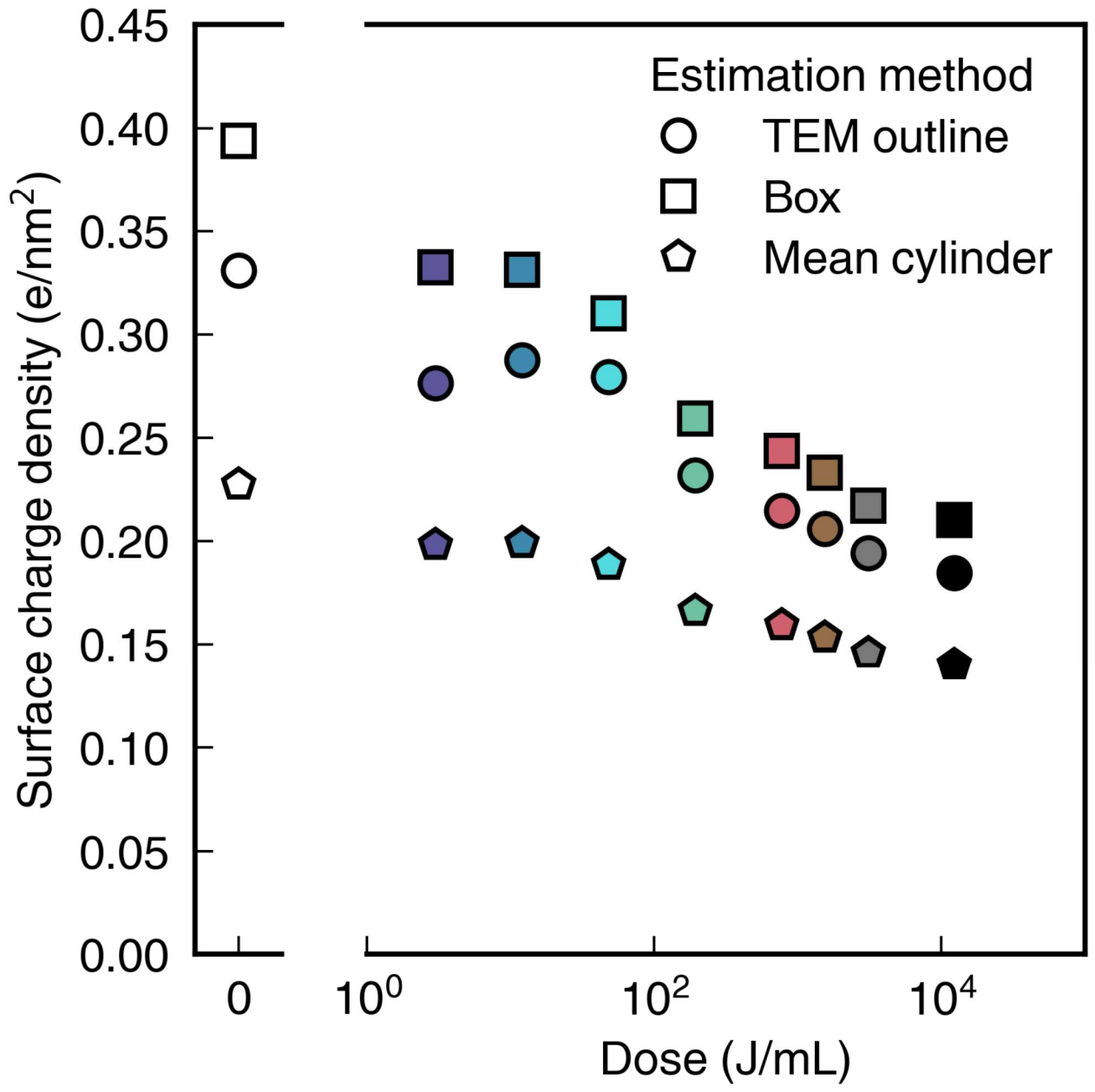

**Figure S4:** Estimates of the CNC surface charge density versus sonication dose. Estimates are based on true particle shape (circles), box properties (squares) or by assuming all CNCs are cylinders with identical length and diameter (pentagons). Estimation methods are described further in the text.

The decrease in CNC surface charge density shown in fig. S4 may be expected to indicate a decrease in colloidal stability. However, a complementary measurement of the mean electrophoretic mobility of the particles was performed by measuring the zeta potential (Section S1.6), which showed no clear trend with sonication dose within the uncertainty of the measurements. This result suggests that the expected decrease in charge per surface area does not make the CNCs colloidally unstable, at least under the conditions of the zeta potential measurement (ionic strength $I \approx 1\,\mathrm{mM}$). However, the decrease in $\langle \sigma \rangle$ with sonication may explain the earlier onset of kinetic arrest at higher doses.

| Sonication dose (J/mL) | Specific surface charge (mmol/kg) | Zeta potential (mV) |
|---:|---:|---:|
| 0 | 154.1 | -37.1 ± 7.5 |
| 12 | 149.9 | -42.0 ± 10.6 |
| 48 | 155.0 | -40.6 ± 10.7 |
| 193 | 154.2 | -41.0 ± 11.7 |
| 772 | 154.3 | -42.8 ± 11.4 |
| 3087 | 152.7 | -41.0 ± 10.8 |

**Table S3:** Table of colloidal properties of CNC suspensions.

# S4 Pitch measurement by SEM on CNC film cross-sections

The pitch of CNC photonic films can be directly observed from the film cross-section in SEM images (Section S1.4). The helicoidal structure of CNC photonic films results in a periodic texture in SEM cross-sections, as shown for selected samples in Figure S3. The pitch is determined from the vertical distance for one full rotation of the structure (corresponding to two repeats of the fingerprint pattern). Cross-sectional SEM images for $u_S$ = 0 J mL$^{-1}$ and $u_S$ >2×10$^3$ J mL$^{-1}$ appeared isotropic with no helicoidal texture.

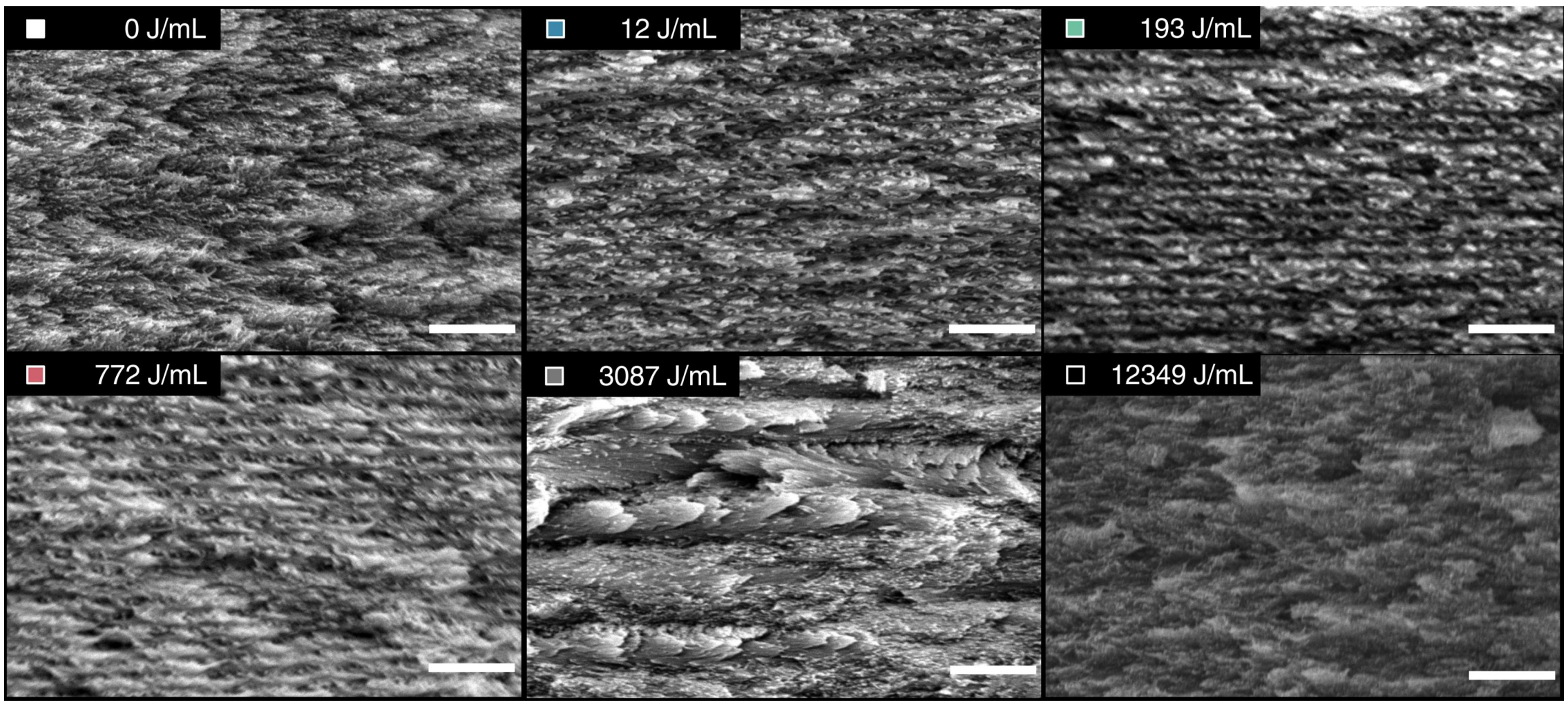

**Figure S5:** Examples of SEM cross-sections for CNC photonic films. Dose indicated by the value on each image. Scale bar is 1 µm.

# S5 Pitch measurement by optical spectroscopy on CNC photonic films

The left-handed helicoidal configuration of birefringent CNCs in a solid film results in selective reflection of left-circular polarised (LCP) light in a wavelength range determined by the cholesteric pitch $P$. The peak reflection wavelength at normal incidence $\lambda_{max}$ is given by

$$\lambda_{max} = n_{\mathrm{CNC}} P, \tag{3}$$

where $n_{\mathrm{CNC}}$ is the average refractive index of the CNC film, which we assumed to be 1.555 based on previous reports (11).

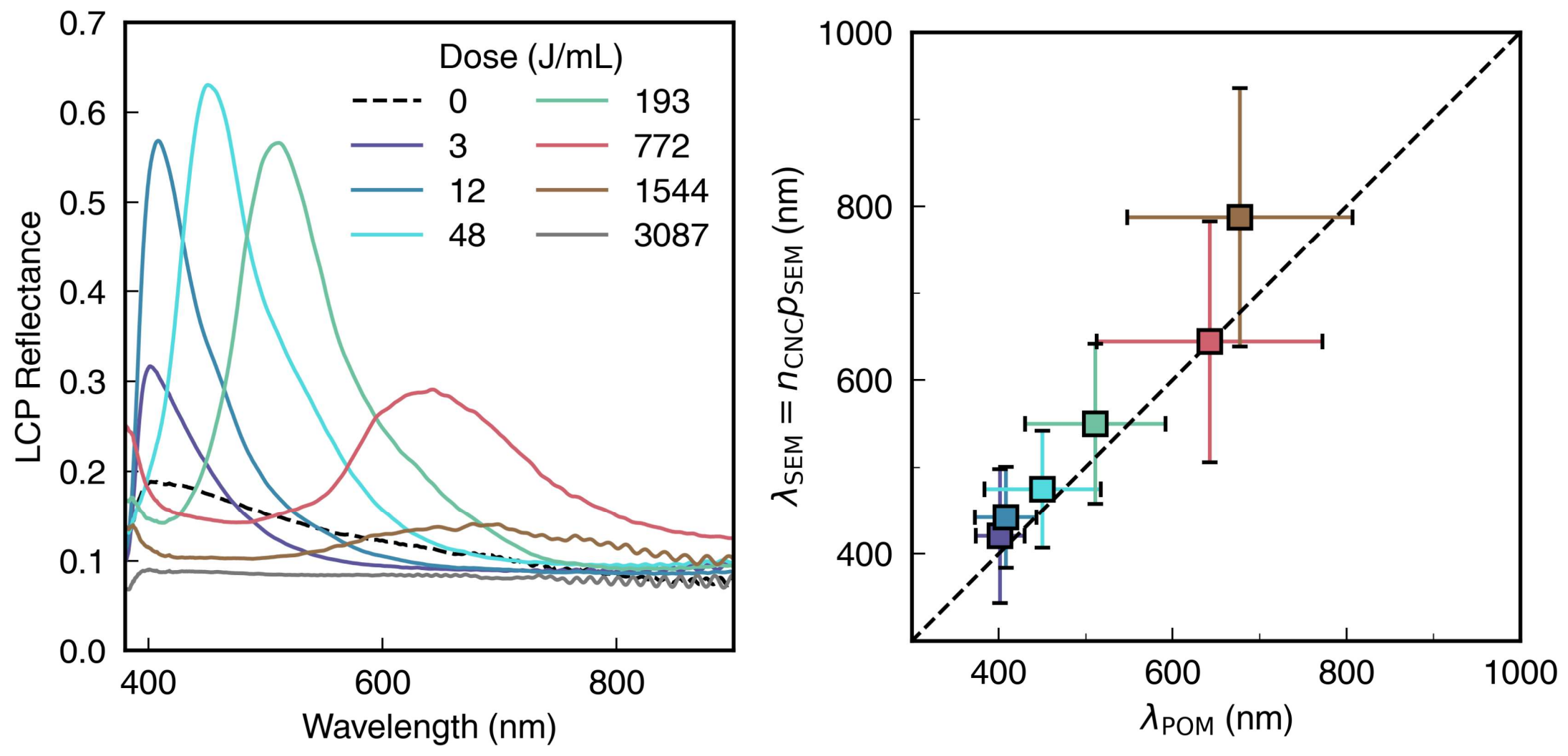

**Figure S6:** (**A**) Left-circular polarised (LCP) reflectance spectra of CNC photonic films. (**B**) Comparison of peak LCP reflectance wavelength from polarised optical microscopy $\lambda_{POM}$ and the predicted peak reflection wavelength based on SEM pitch values $\lambda_{SEM} = n_{CNC}P_{SEM}$, assuming $n_{CNC} = 1.555$. The error bars for $\lambda_{POM}$ indicate the peak full width at half maximum, while the error bars for $\lambda_{SEM}$ indicate the standard deviation of the pitch measurements.

Left-circular polarised reflection spectra were obtained at a range of sonication doses (Section S1.3). The peak reflection wavelength increases with sonication dose, as shown in Figure S6. The apparent peak for the never-sonicated sample ($u_S = 0\,\mathrm{J\,mL^{-1}}$) could indicate the presence of some chiral nematic domains within the structure, although no ordering was visible in SEM images of the film cross-section (fig. S5). This small peak could also be attributed to scattering in the disordered structure, or partial absorption by the reference silver mirror in this wavelength range. The peak reflection wavelength at high doses ($u_S > 2000\,\mathrm{J\,mL^{-1}}$) was too far into the infrared to be detected using this setup. There is reasonably good agreement between the peak LCP reflectance wavelength obtained by spectroscopy and the pitches measured by SEM (Fig. 1c). Determination of CNC pitch by SEM tends to over-estimate the true value as the pitch of tilted chiral nematic domains appears larger when viewed from the plane of film breakage.

# S6 Evidence of bundles from cryoTEM imaging

The preparation of CNC samples for conventional TEM imaging involves drying a droplet of sample onto the TEM grid, which can cause aggregates to appear in TEM images that do not exist in the original suspension. These artefacts should not be confused with CNC bundles, which are native to CNC suspensions at all sonication doses, or with large clusters of bundles at low sonication doses (classified as aggregates in the article). To illustrate that bundles are a native feature of CNC suspensions, cryoTEM imaging was performed on selected CNC samples (Section S1.8). In cryoTEM imaging, the CNC suspension is frozen and never dried, thus eliminating any risk of artefact creation. Example images are shown in Figure S7. It is evident from qualitative inspection of the images that these suspensions contain CNCs with a bundled morphology.

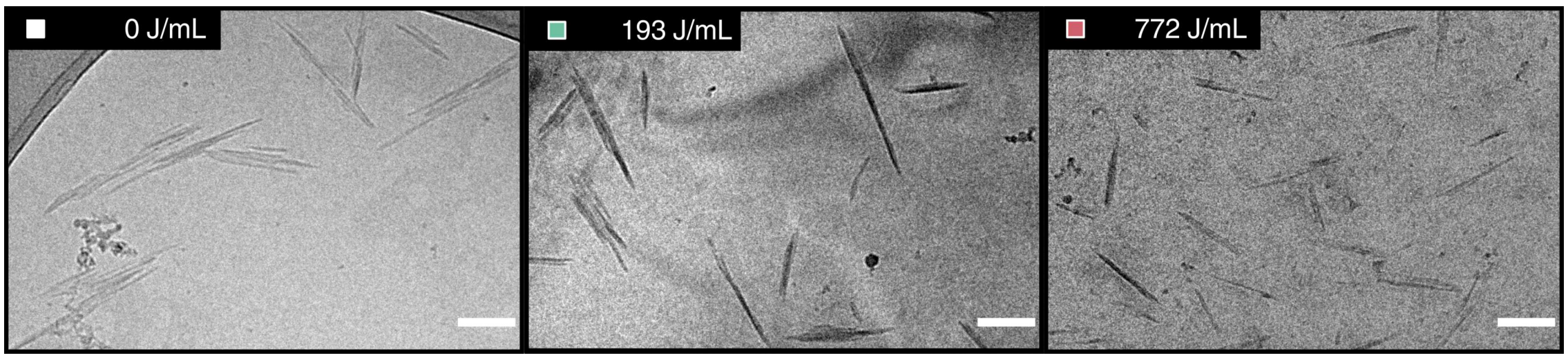

**Figure S7:** Examples of cryoTEM images for CNC suspension, illustrating the presence of bundles. Sonication dose $u_S$ is indicated by the value on each each image. Scale bar is 200 nm.

# S7 Definitions and distributions of morphological properties

The morphological properties used in this work are summarised in tables S4 to S6. The corresponding distributions for each sonication dose are figs. S8 to S10

| Property | Symbol | Equation | Units |
|---|---|---|---|
| Area, true shape | $A$ | | $nm^2$ |
| Area, convex hull | $A_{CH}$ | | $nm^2$ |
| Area, oriented bounding box | $A_b$ | $L_b W_b$ | $nm^2$ |
| Perimeter, true shape | $P$ | | nm |
| Perimeter, convex hull | $P_{CH}$ | | nm |
| Length, oriented bounding box | $L_b$ | | nm |
| Length, Feret (max. caliper length) | $L_F$ | | nm |
| Width, oriented bounding box | $W_b$ | | nm |
| Width, area-equivalent (AE) | $W_{AE}$ | $A/L_b = \mathcal{R} W_b$ | nm |
| Width, Feret (min. caliper length) | $W_F$ | | nm |

**Table S4:** Table of size properties measured from TEM images, with definitions. The distributions for $L_b$ and $W_{AE}$ are given in the main text; the distributions for other properties are plotted in fig. S8.

| Property | Symbol | Equation | Units |
|---|---|---|---|
| Aspect ratio, oriented bounding box | $\alpha_b$ | $L_b/W_b$ | - |
| Aspect ratio, area-equivalent (AE) | $\alpha_{AE}$ | $L_b/W_{AE} = L_b^2/A$ | - |
| Isoperimetric Quotient, 2D (Thinness Ratio) | $IQ_2$ | $4\pi A/P^2$ | - |
| Solidity | $\mathcal{S}$ | $A/A_{CH}$ | - |
| Convexity | $\mathcal{C}$ | $P/P_{CH}$ | - |
| Rectangularity | $\mathcal{R}$ | $A/A_b$ | - |
| Perimetric rectangularity | $\mathcal{R}_P$ | $P/2(L_b + W_b)$ | - |

**Table S5:** Table of shape properties measured from TEM images, with definitions. Distributions for $\mathcal{R}$ are given in the main text; the distributions for other properties are plotted in fig. S9.

| Property | Symbol | Equation | Units |
|---|---|---|---|
| Thickness, particle mean | $\langle T \rangle$ | | nm |
| Area, particle surface | $\Sigma$ | $2A + P\langle T \rangle$ | $nm^2$ |
| Volume | $V$ | $A\langle T \rangle$ | $nm^3$ |
| Specific surface area | – | $\Sigma/(\rho_{CNC} V)$ | $m^2/g$ |
| Isoperimetric Quotient, 3D | $IQ_3$ | $36\pi V^2/\Sigma^3$ | - |
| Circle-equivalent (CE) diameter | $D_{CE}$ | $\sqrt{4 W_{AE}\langle T \rangle/\pi}$ | - |
| Aspect ratio, 3D | $\alpha_{3D}$ | $L_b/\sqrt{W_{AE}\langle T \rangle}$ | nm |
| Cross-section (XS) aspect ratio, area equivalent | $\alpha_{XS,AE}$ | $W_{AE}/\langle T \rangle$ | - |
| Cross-section (XS) aspect ratio, box | $\alpha_{XS,b}$ | $W_b/\langle T \rangle$ | - |
| Max-min (MM) aspect ratio, box | $\alpha_{MM}$ | $L_b/\langle T \rangle$ | - |

**Table S6:** Table of size and shape properties estimated using AFM and TEM data, with definitions. Distributions for $\langle T \rangle$ are plotted in fig. S12. Distributions for $\alpha_{3D}$ are given in the main text. Distributions for other properties are plotted in fig. S10.

| $u_S$ (J/mL) | # particles, TEM |
|---|---|
| 0 | 254 |
| 3 | 258 |
| 12 | 257 |
| 48 | 263 |
| 193 | 256 |
| 772 | 396 |
| 1544 | 506 |
| 3087 | 485 |
| 12349 | 540 |

**Table S7:** Table of counts of CNC particles at each sonication dose measured in TEM morphological analysis.

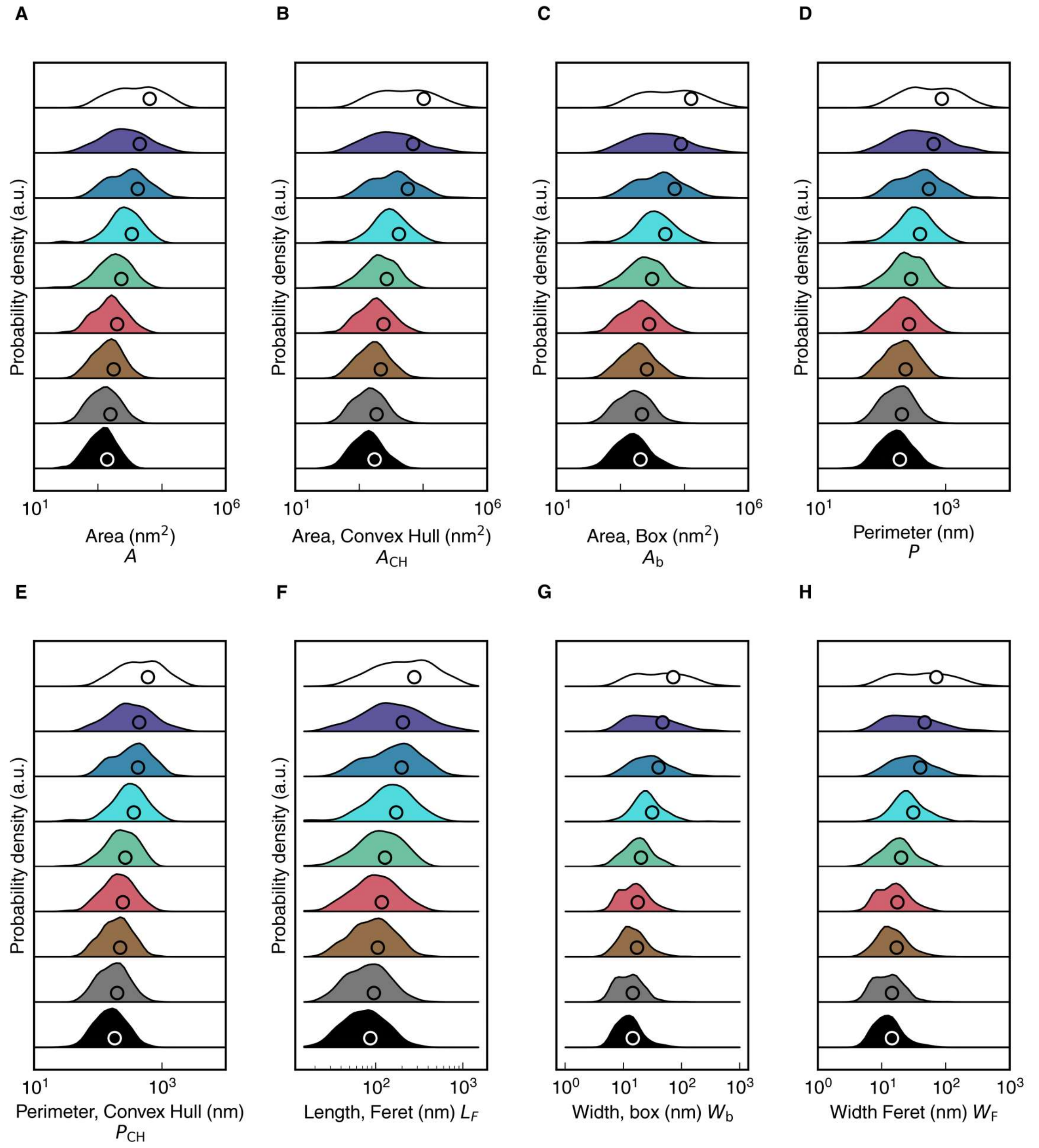

**Figure S8:** Size distributions for additional properties: (**A**) Area $A$ (**B**) Area of convex hull $A_{\mathrm{CH}}$ (**C**) Area of oriented bounding box $A_{\mathrm{b}}$ (**D**) Perimeter $P$ (**E**) Perimeter of convex hull $P_{\mathrm{CH}}$ (**F**) Feret length $L_F$ (**G**) Width of oriented bounding box $W_{\mathrm{b}}$ and (**H**) Feret width (minimum caliper length) $W_F$ The properties are defined in Table S5.

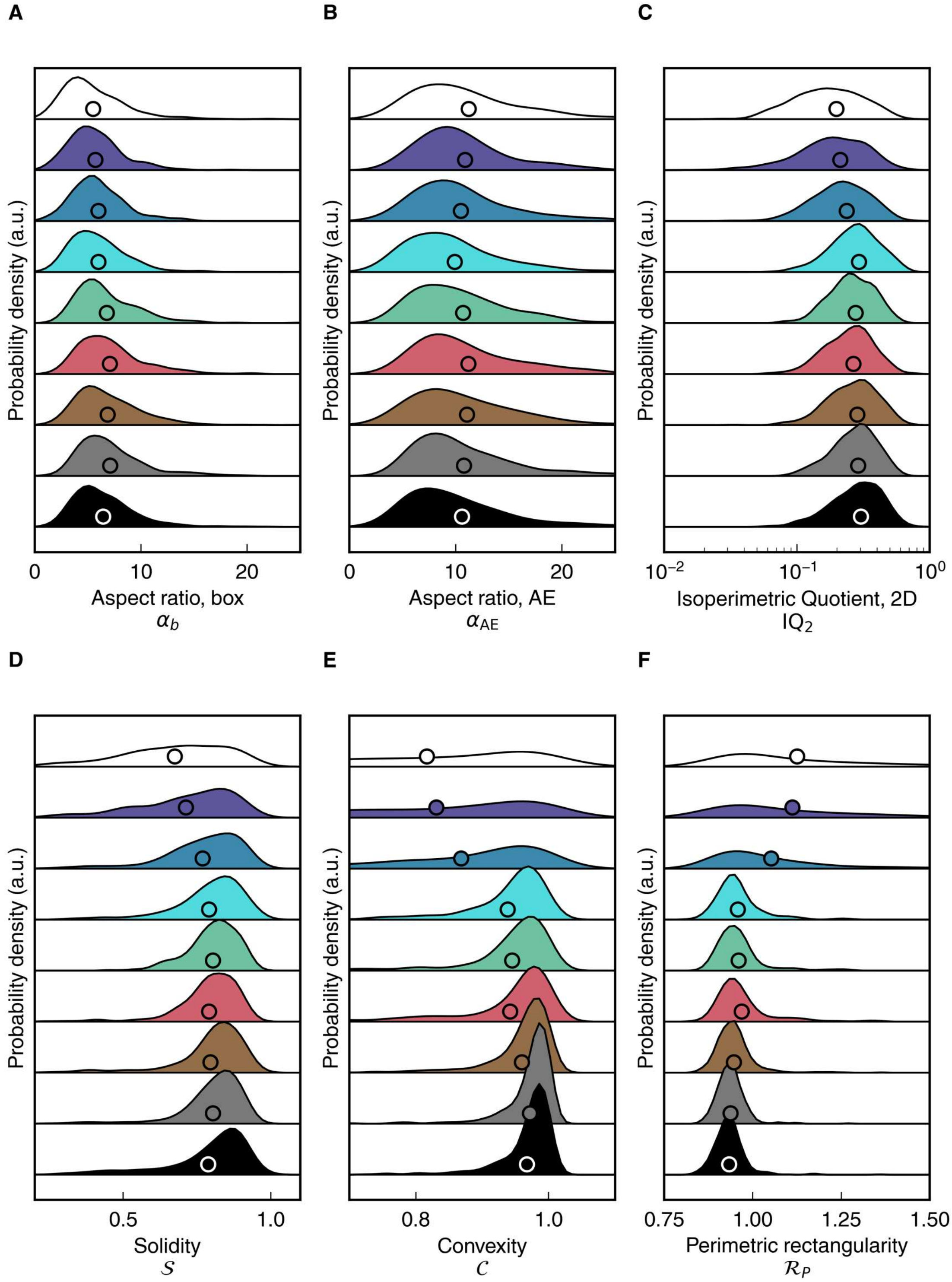

**Figure S9:** Shape distributions for additional properties: (**A**) Aspect ratio of oriented bounding box $\alpha_b$ (**B**) Aspect ratio using box length and area-equivalent width $\alpha_{AE}$ (**C**) 2D Isoperimetric quotient $IQ_2$ (**D**) Solidity $\mathcal{S}$ (**E**) Convexity $\mathcal{C}$ (**F**) Perimetric rectangularity $\mathcal{R}_P$.The properties are defined in Table S5.

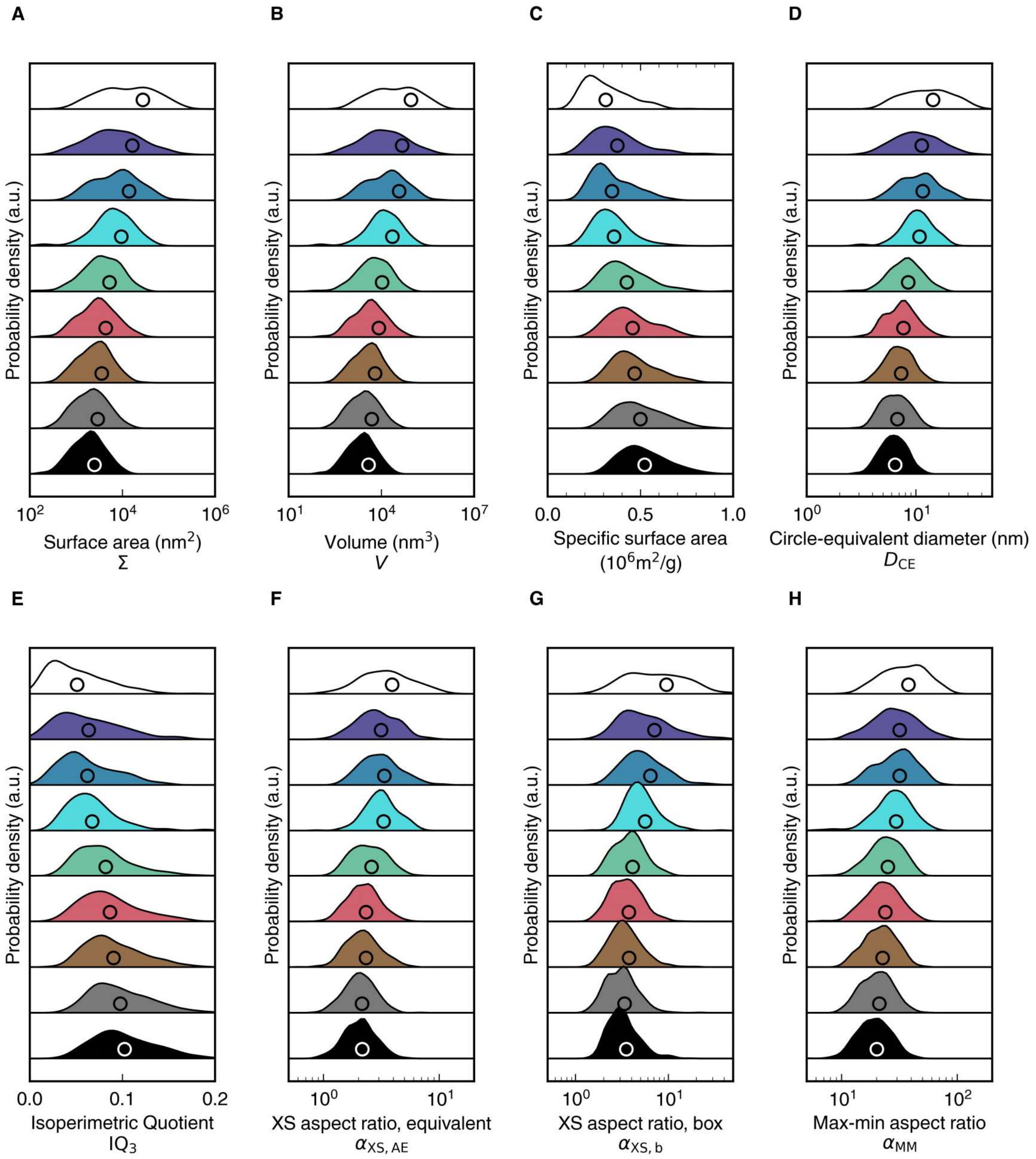

**Figure S10:** Shape distributions for additional 3D morphological properties: (**A**) Surface area $\Sigma$ (**B**) Volume $V$ (**C**) Specific surface area (**D**) circle-equivalent diameter $D_{CE}$ (**E**) 3D Isoperimetric quotient $IQ_3$ (**F**) Cross-sectional aspect ratio using AE width $\alpha_{XS,AE}$ (**G**) Cross-sectional aspect ratio using box width $\alpha_{XS,b}$ (**H**) Max-min aspect ratio $\alpha_{MM}$. These properties are defined in Table S6.

# S8 Estimation of CNC thickness using atomic force microscopy (AFM)

Atomic force microscopy (AFM) provides data on the topography of a sample surface, and therefore offers complementary morphological information to transmission electron microscopy (TEM). For CNCs deposited on a flat substrate and imaged using AFM, the mean and maxi-mum thickness of the particle can be accurately measured, as illustrated in Figure S11A. The lateral size of particles, as expressed as the Feret length and width (Figure S11B), can also be obtained. However, it should be noted that the $XY$ topography of the particle is convolved by the effective diameter of the AFM tip (Figure S11C), which limits the use of AFM data for more detailed morphological analysis.

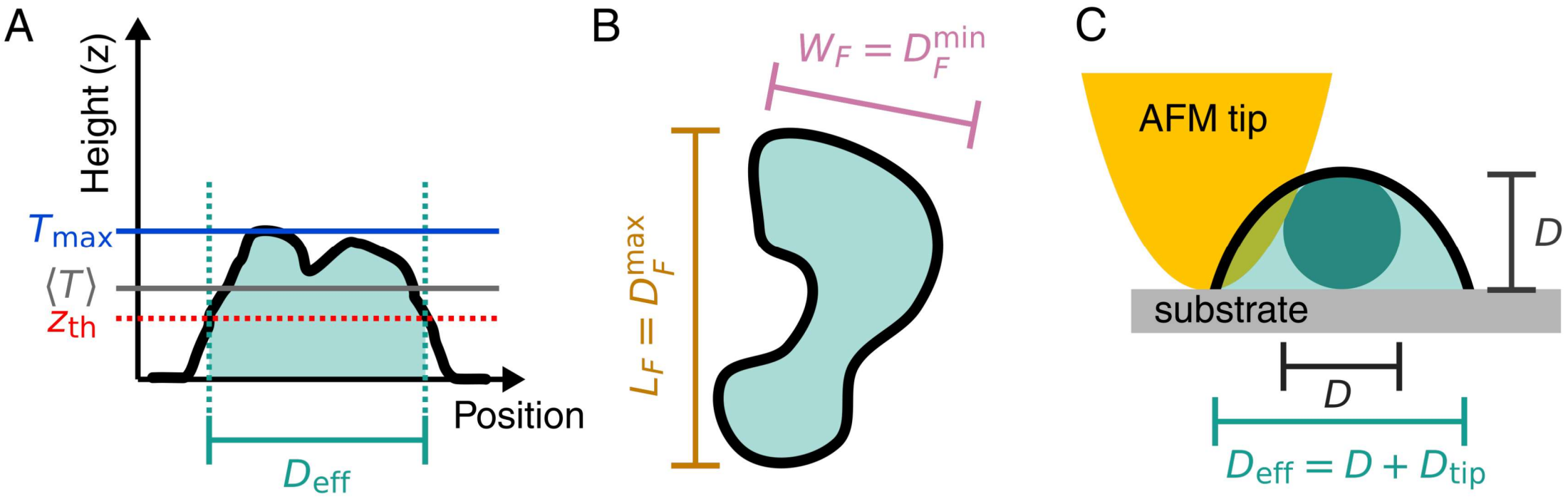

**Figure S11:** Morphological properties obtained from atomic force microscopy (AFM). (**A**) Individual particles (also known as "grains" in AFM terminology) are identified as the region where the baseline-corrected surface height $z(x, y)$ exceeds a critical threshold value $z_{th}$ (red dotted line). Each particle is defined as the bounded region that satisfies the threshold criterion $z \geq z_{th}$. A threshold of $z_{th}$ = 1.2 nm was used in this work. The mean thickness $\langle T \rangle$ of each particle (grey line) is obtained by averaging the height value within the particle boundary. The maximum thickness $T_{max}$ (blue line) can also be obtained. (**B**) The Feret (caliper) diameter of a particle in the substrate plane, $D_F$, is defined as the smallest distance between oriented parallel lines that encloses the particle. The value of $D_F$ depends on the orientation of the parallel lines relative to the particle. The Feret length $L_F$ (orange) and Feret width $W_F$ (pink) of the particle are defined as the maximum and minimum Feret diameter respectively. Note that the parallel lines that enclose $L_F$ and $W_F$ are not at perpendicular orientations (unlike the definition of the box length $L_b$ and box width $W_b$, table S5). (**C**) Illustration of the effect of a tip artefact on a spherical nanoparticle (green) on a flat substrate. For a spherical nanoparticle, the bare particle diameter $D$ is accurately measured in the height profile but not in the lateral scan direction due to the width of the AFM tip $D_{tip}$, leading to an effective diameter $D_{eff}$ being measured.

CNCs at a range of sonication doses were imaged and analysed using AFM (see Section S1.9 for the experimental method and table S8 for the number of particles measured for each dose). This analysis shows that sonication reduces the thickness of CNC particles (Figure S12), as well as the length and width (Figure S13).

| $u_S$ (J/mL) | # particles, AFM |
|---|---|
| 0 | 90 |
| 12 | 302 |
| 193 | 369 |
| 772 | 554 |
| 12349 | 1344 |

**Table S8:** Table of counts of CNC particles at a range of sonication doses measured in AFM morphological analysis.

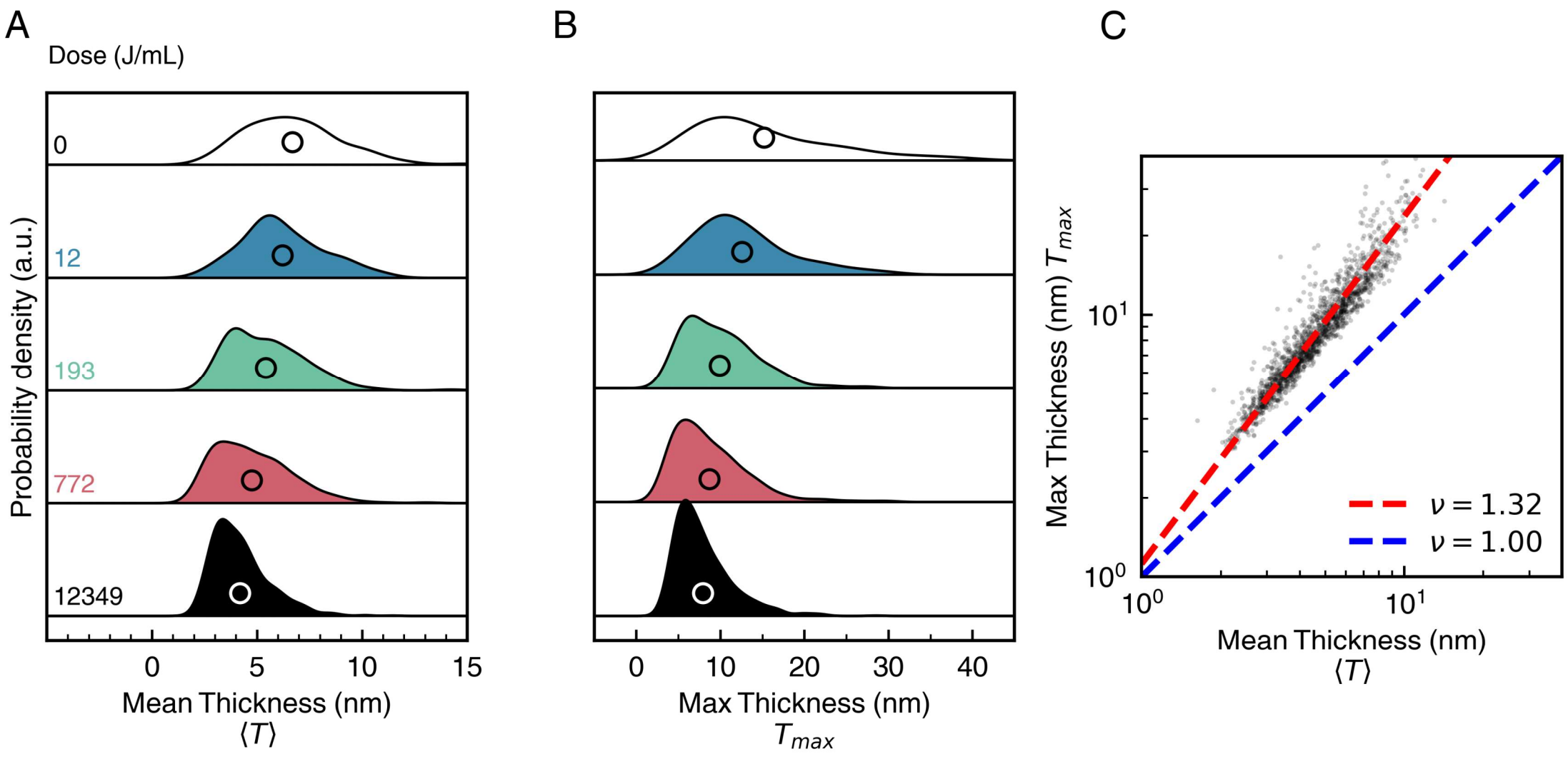

**Figure S12:** Statistics for CNC thickness from AFM image analysis. Size distributions for mean thickness $\langle T \rangle$ (**A**) and max thickness $T_{\max}$ (**B**), demonstrating the reduction in thickness with sonication dose limited by the minimum thickness of a single cellulose I$\beta$ crystallite ($\approx$ 3 nm). (**C**) 2D scatter plot of mean versus max thickness, combining data for all sonication doses. Remarkably, a power law relation between mean and max thickness ($T_{\max} \propto \langle T \rangle^{\gamma}$) is well-described by an exponent $\nu = 1.32 \approx 4/3$ (red dotted line), rather than a linear fitting ($\nu = 1$, blue dotted line, shifted vertically for clarity). This scaling behaviour can be attributed to the bundled, fractal-like 3D morphology of the CNC particles. It should be noted that different particle populations (Aggregates, Bundles, Crystallites and Distorted crystallites) are expected to show different scaling behaviour, but the classification used for particle shapes from TEM images could not be directly applied to AFM shape data.

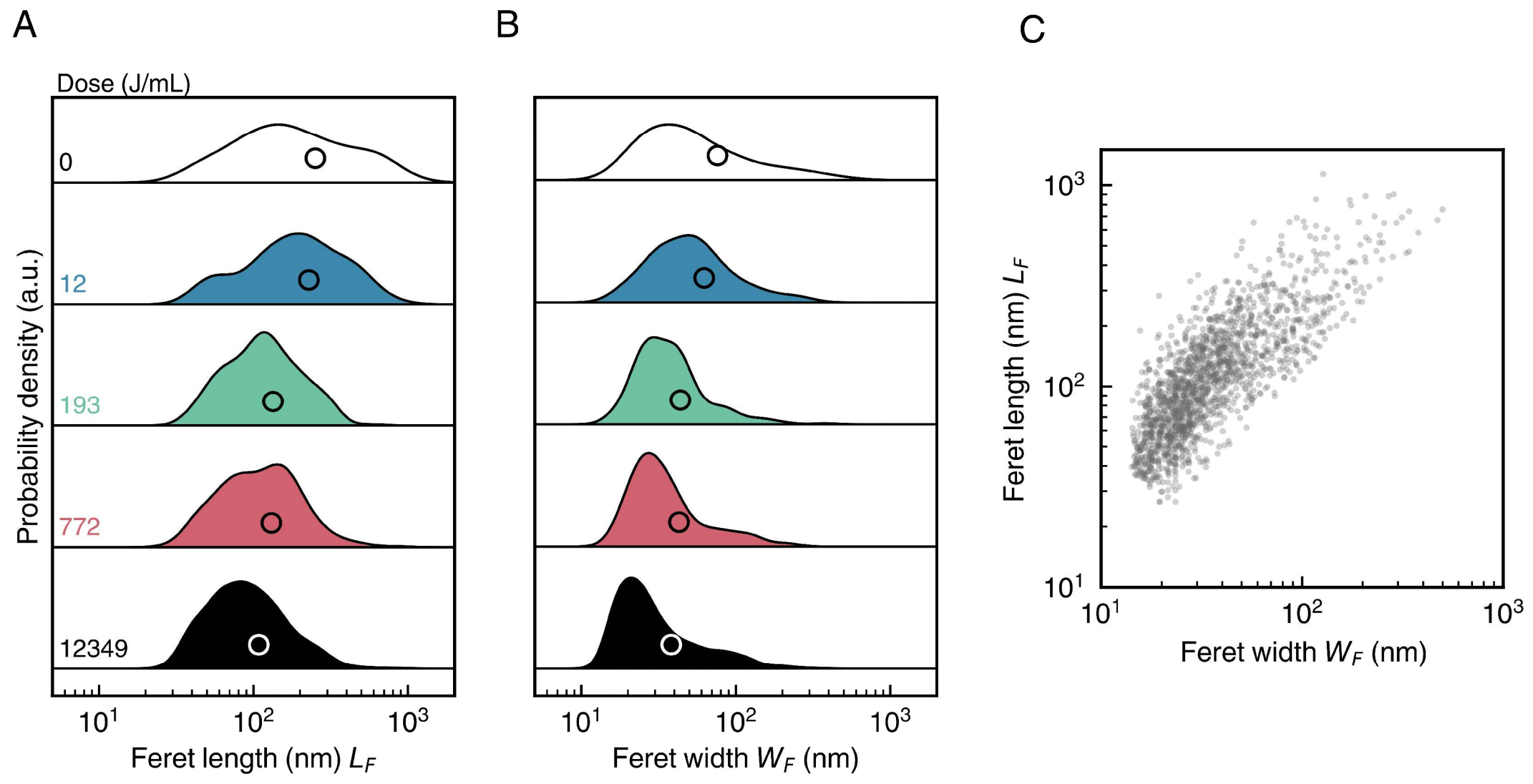

**Figure S13:** Statistics for Feret length and width from AFM image analysis. Size distributions for Feret length $L_F$ (**A**) and Feret width $W_F$ (**B**), demonstrating the reduction in overall particle size with increasing sonication dose. Circles indicate mean values. (**C**) 2D scatter plot of Feret width versus Feret length, combining data for all sonication doses.

AFM data can also be used to estimate the thickness of a given CNC from a TEM image by matching the particle length to a length-thickness calibration curve obtained from AFM image analysis. As shown in Figure S14A, the AFM Feret length shows a clear positive correlation with mean thickness, which can be modeled by an empirical power law relation. Length values from AFM and TEM show fairly good agreement, even without correcting for the tip diameter (Figure S14B). Consequently, the thickness of particles measured from TEM images was estimated by assuming $L_F^{\mathrm{TEM}} = L_F^{\mathrm{AFM}}$ and applying the power law fitting from Figure S14A. These thickness values were then used to estimate other shape properties (e.g. the 3D aspect ratio $\alpha_{\mathrm{3D}}$).

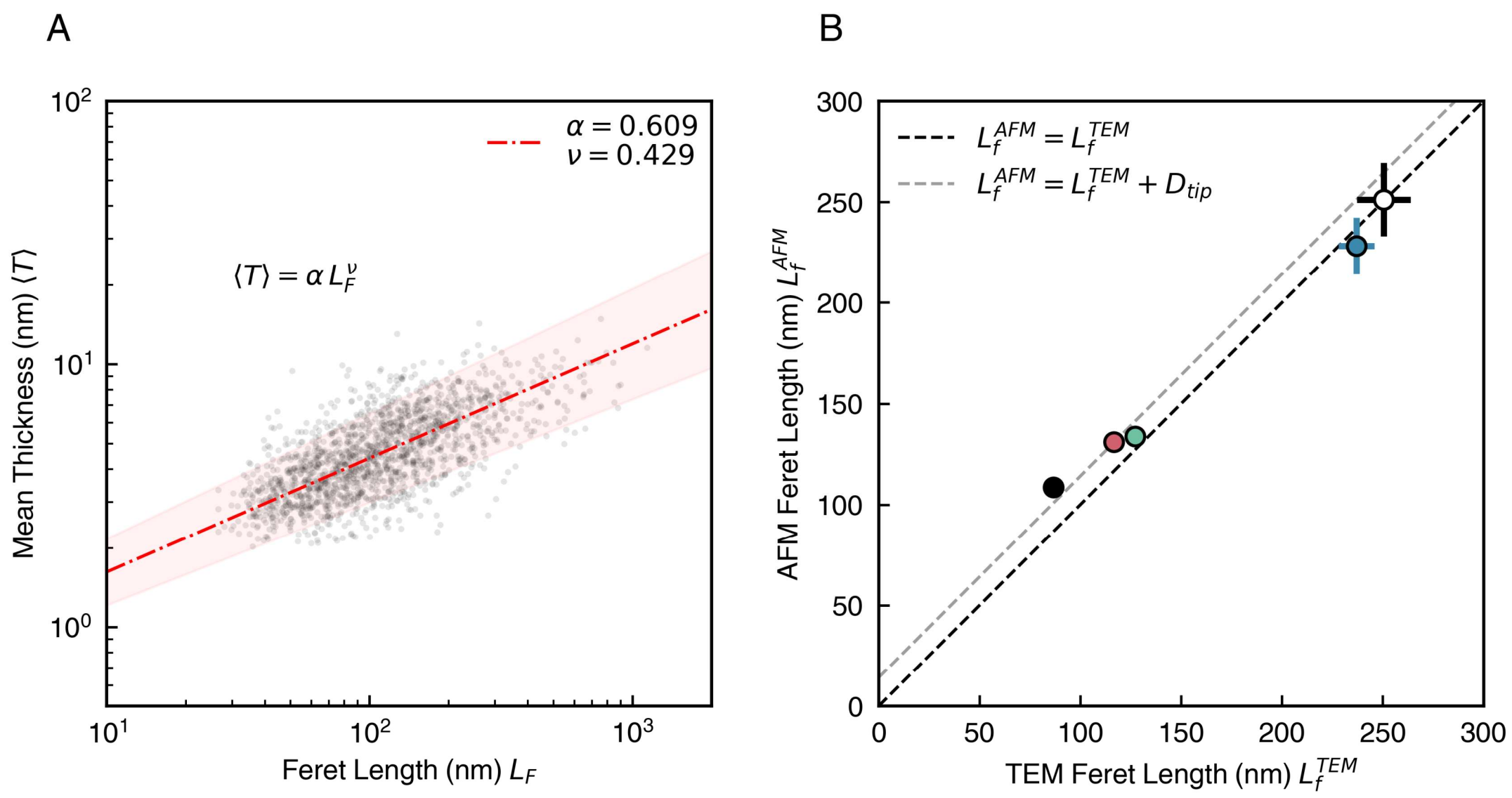

**Figure S14:** (**A**) Correlation between Feret length $L_F$ and mean thickness $\langle T \rangle$ from AFM image analysis. Red line indicates power law fitting $\langle T \rangle = AL_F^{\nu}$. Red shaded area indicates uncertainty. (**B**) Comparison of mean Feret length obtained from TEM analysis ($L_F^{\mathrm{TEM}}$) versus AFM ($L_F^{\mathrm{AFM}}$). Error bars indicate standard error of the mean. Good agreement is obtained between TEM and AFM values when adding a correction for the tip diameter (grey dotted line), especially at high sonication doses where the error bars are smaller.

# S9 Estimation of ensemble size values from individual particle sizes

The size distributions obtained from TEM images can be used to estimate ensemble properties such as the mean particle cross-section from UV-vis transmission spectroscopy (section S1.7, section S12) or the mean hydrodynamic diameter obtained from DLS measurements (section S1.1). The mean particle cross-section was estimated using eq. (23) in section S12, while the hydrodynamic diameter was estimated by the method explained below.

The mean hydrodynamic diameter $\langle d_h \rangle$ obtained from a DLS measurement is the hydrodynamic diameter corresponding to the average translational diffusion coefficient $\langle D_t \rangle$:

$$\langle d_h \rangle = \frac{k_B T}{3\pi\eta_0 \langle D_t \rangle} \tag{4}$$

where $k_B$ is the Boltzmann constant, $T$ is temperature, $\eta_0$ is the suspension viscosity (assumed to be the viscosity of water).

The average diffusion coefficient $\langle D_t \rangle$ is weighted by the scattered light intensity, which scales with the particle volume squared. The particle sizes measured from TEM can be used to estimate the intensity-averaged diffusion coefficient:

$$\langle D_t \rangle = \frac{\sum_p v_p^2 D_t}{\sum_p v_p^2} \tag{5}$$

where $v_p$ is the particle volume and $\sum_p$ indicates summation of all the particles measured. The translation diffusion coefficient of CNCs can be estimated by assuming CNCs diffuse like ideal rods. The diffusion coefficient for an ideal rod of length $L$ and aspect ratio $\alpha$ is (12):

$$D_t = \frac{k_B T}{3\pi\eta_0} \frac{\ln\alpha + \nu(\alpha)}{L} \tag{6}$$

where $\nu(\alpha)$ is a correction added to account for end effects:

$$\nu(\alpha) \approx 0.312 + 0.565\frac{1}{\alpha} - 0.100\frac{1}{\alpha^2}. \tag{7}$$

To apply these equations to CNC suspensions, the particle length was assumed to be the box length ($L_{\text{eff}} = L_b$). The effective aspect ratio was determined by first calculating the effective diameter of the CNC cross-section:

$$D_{\text{eff}} = \sqrt{W_{\text{AE}} \langle T \rangle} \tag{8}$$

where $W_{\text{AE}}$ is the AE width discussed in the article, and $\langle T \rangle$ is the estimated particle thickness (Section S8). The effective aspect ratio can then be calculated as

$$\alpha_{\text{eff}} = \frac{L_{\text{eff}}}{D_{\text{eff}}} \tag{9}$$

The effective hydrodynamic size of charged colloidal particles in suspension can also be modified by electroviscous effects due to the electric double layer (13). For the CNC suspensions used for DLS measurements in this work, the hydrodynamic diameter is $d$ ~100 nm and the Debye length $\kappa^{-1}$ ~ 10 nm. In this limit ($\kappa d$ ~ 10), electroviscous effects are expected to be negligible (<5%), and are therefore not taken into account (13).

The estimated $z$-average hydrodynamic diameter calculated using Equations (4) to (7) is shown in Figure S15A alongside the experimental values from DLS. The particle cross-section was estimated using eq. (23) and is shown in Figure S15B alongside values obtained from UV-vis transmission spectroscopy (Section S12). The estimated values capture the overall trends and show good agreement, especially at higher sonication doses. The discrepancies at low sonication dose can be attributed to the irregular shapes of large particles, which deviate from the cylinder model, and also the sensitivity of the scattering intensity to fluctuations in the number of large particles observed in TEM images.

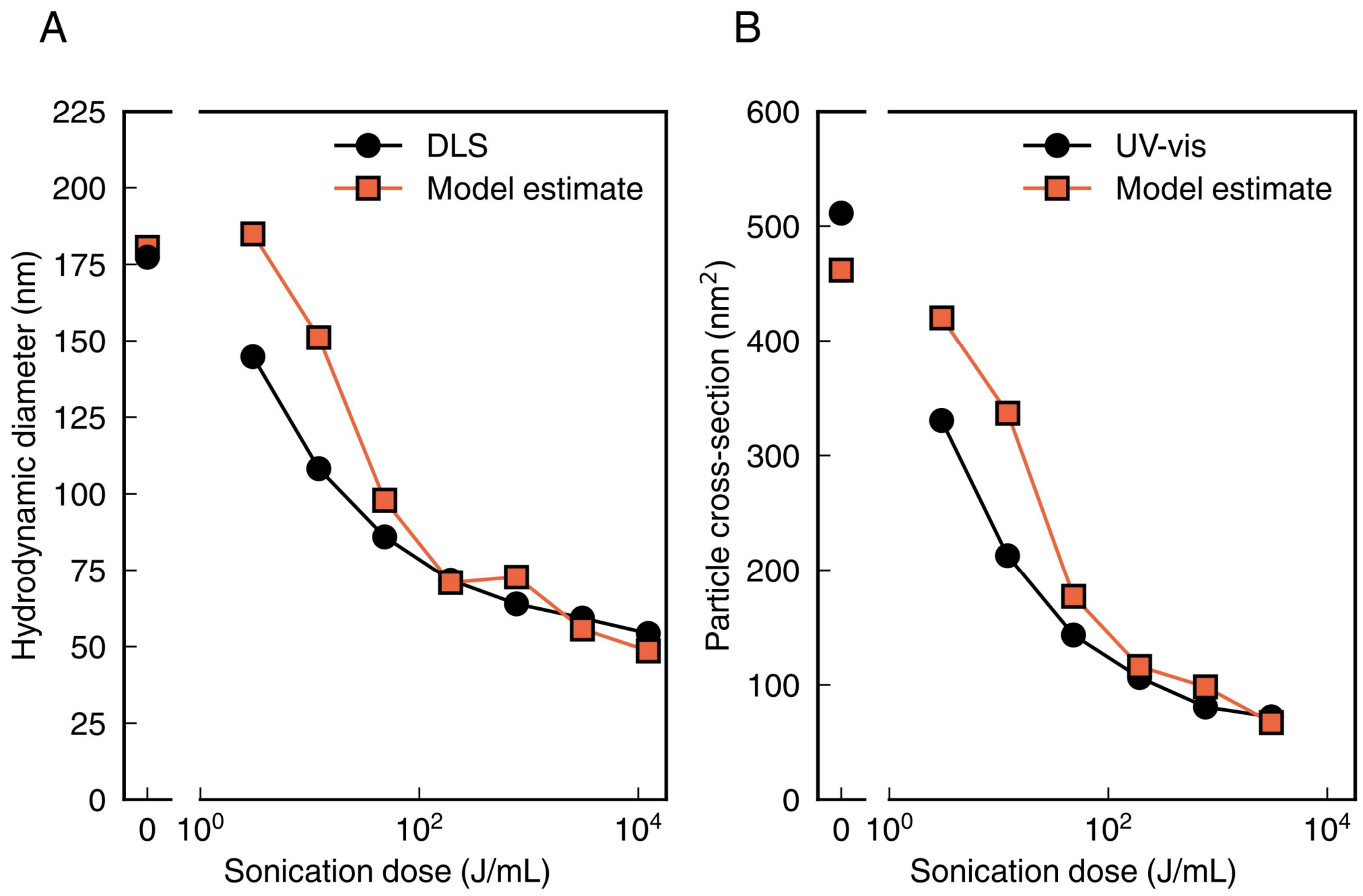

**Figure S15:** Comparison of experimentally-measured ensemble size properties and estimated values based on individual particle sizes in TEM images. (**A**) Hydrodynamic diameter (**B**) Particle cross-section.

# S10 Correlating CNC phase behaviour with particle morphology

For cylindrical particles forming a lyotropic liquid crystal phase, the concentrations of the isotropic-biphasic and biphasic-anisotropic phase boundaries ($\phi_0$ and $\phi_1$ respectively), and the position of the midpoint of the biphasic region ($\phi_m = (\phi_0 + \phi_1)/2$), are expected to be inversely proportional to the cylinder aspect ratio, as shown by Onsager and subsequent authors (14; 15).

In this work, the biphasic midpoint $\phi_m$ is reasonably well-described by a linear fitting with inverse aspect ratio (fig. S16A), in agreement with the theoretical predictions. However, only $\phi_1$ is observed to increase with sonication dose, while $\phi_0$ appears to remain constant. In terms of re-scaled volume fraction $\phi\alpha$, both $\phi_1\alpha$ diverges with sonication dose (fig. S16B). This broadening can be attributed to an increase in the coefficient of variation (relative polydispersity) in aspect ratio $\tilde{\sigma}_\alpha = \sqrt{(\langle\alpha^2\rangle - \langle\alpha\rangle^2)/\langle\alpha\rangle^2}$. Figure S16C shows the variation in $\tilde{\sigma}_\alpha$ with re-scaled volume fraction, which shows trends similar to previous theoretical work on length-polydisperse rods (16).

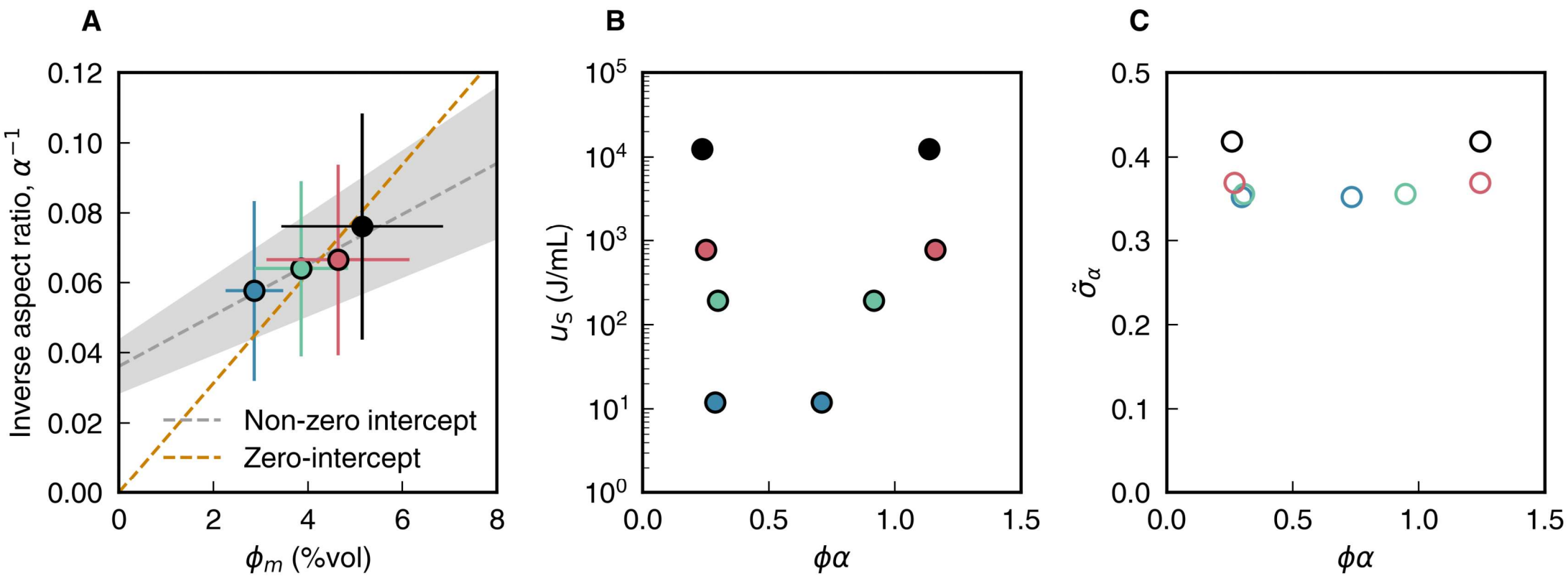

**Figure S16:** (**A**) Correlation between inverse aspect ratio and midpoint of the biphasic region. Error bars in biphasic midpoint indicate inter-quartile range. Grey line indicates linear fitting. Orange line indicates linear fitting forced to pass through the origin. (**B**) Biphasic boundary points ($\phi_0, \phi_1$) plotted versus re-scaled volume fraction $\phi\alpha$. (**C**) Aspect ratio polydispersity $\tilde{\sigma}_\alpha$ plotted versus re-scaled volume fraction $\phi\alpha$.

# S11 Classification of CNC particles

The classification of CNC particles into four classes (A, B, C, D) is based on their shape properties observed in TEM images, as described in the article. The relative number fraction of each particle class ($X = A, B, C, D$) in the overall population is given by

$$\mathcal{N}_X = \frac{N_X}{\sum_X N_X} \tag{10}$$

where $N_X$ is the number of particles in class $X$. The relative number fraction for each sonication dose in shown in fig. S17

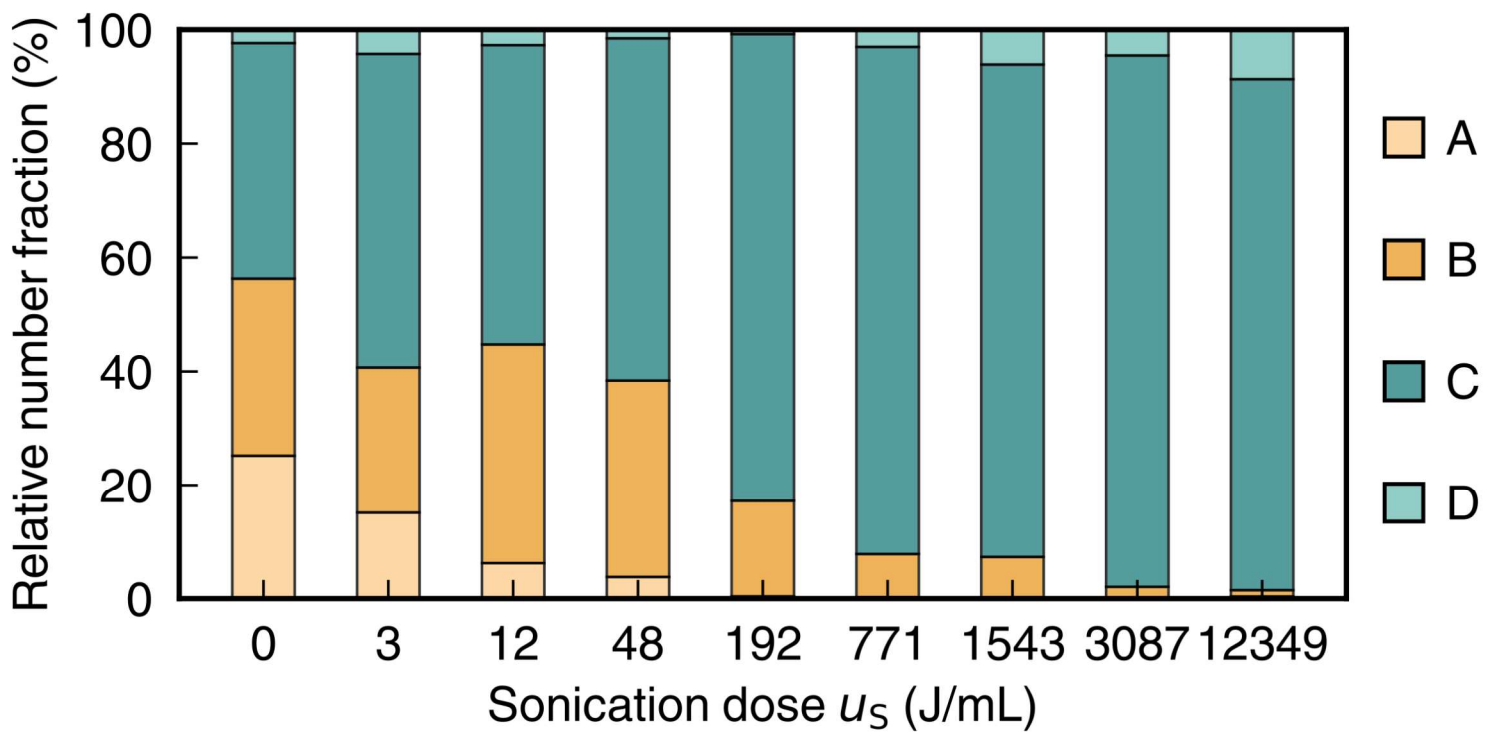

**Figure S17:** Relative number fraction $\mathcal{N}_X$ for each particle class $X = A, B, C, D$.

Alternatively, the relative volume fraction of each particle class can also be calculated

$$\mathcal{V}_X = \frac{V_X}{\sum_X V_X} \tag{11}$$

where $V_X$ is the total volume of all particles in class $X$. To calculate $V_X$, the volume of each CNC in class $X$ (index $j$) is estimated using the expression

$$v_j = A_j \langle T \rangle_j \tag{12}$$

where $A_j$ is the projected area observed in the TEM image and $\langle T \rangle_j$ is the estimated mean particle thickness (see section S8 for details). The total volume is then given by

$$V_X = \sum_{j \in X} v_j \tag{13}$$

Note that the relative volume fractions are normalised such that $\sum_X \mathcal{V}_X = 1$. The absolute volume fraction of particles of class $X$ (in the suspension) is given by $\phi_X = \mathcal{V}_X \phi_{\text{CNC}}$. The relative volume fractions versus sonication dose is shown in the article.

The differences in morphological properties between the sub-populations can be clearly seen in histograms for each population. Histograms for particle box length $L_b$ and 3D aspect ratio $\alpha_{\text{3D}}$ of each class are shown in fig. S18. Histograms for the cross-section aspect ratio expressed as $\alpha_{\text{XS,AE}}$ and $\alpha_{\text{XS,b}}$ (defined in table S6) are shown in fig. S19. Histograms for the 2D isoperimetric quotient and estimated 3D isoperimetric quotient are shown in fig. S20.

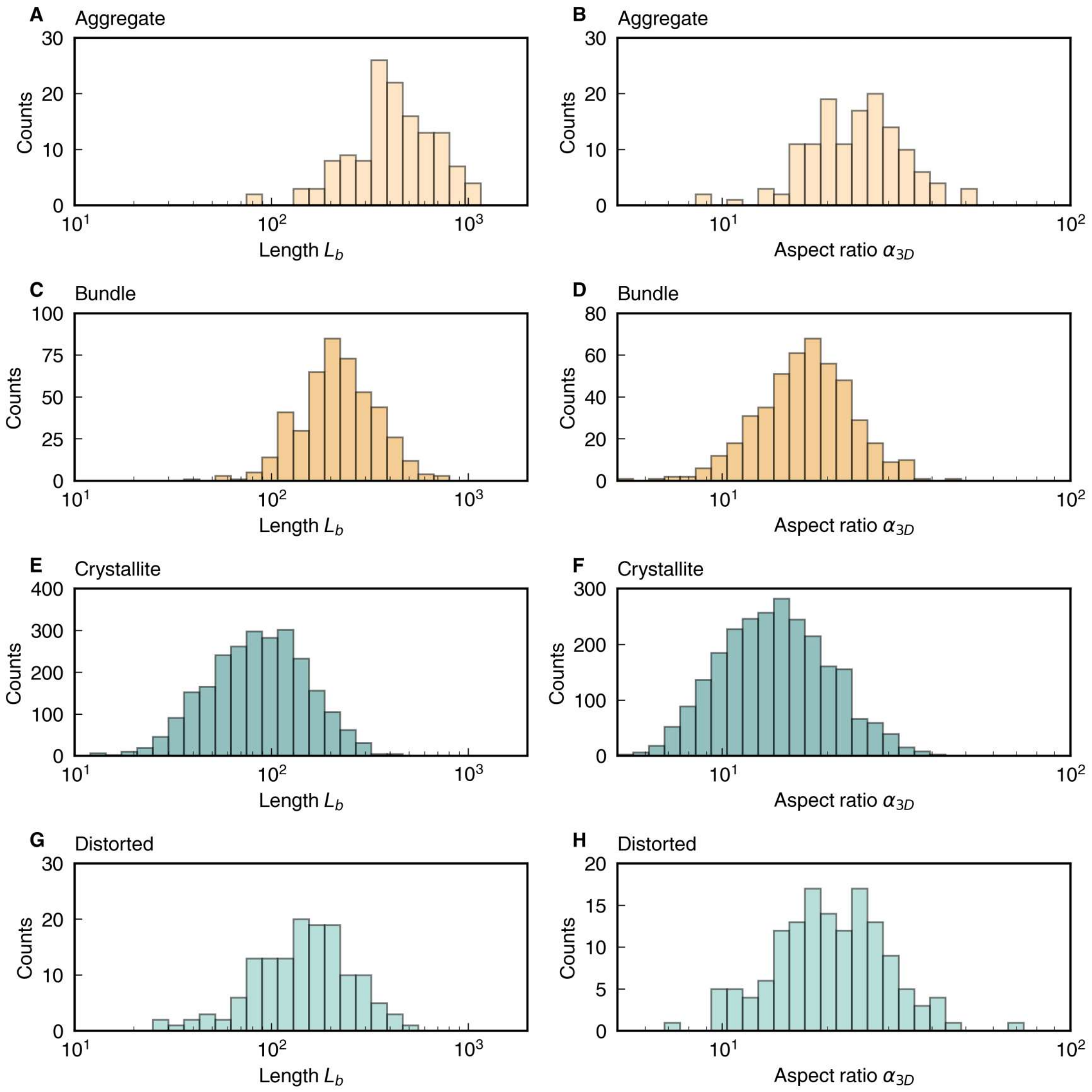

**Figure S18:** Histograms for particle length $L_b$ and 3D aspect ratio $\alpha_{3D}$ for each particle class.

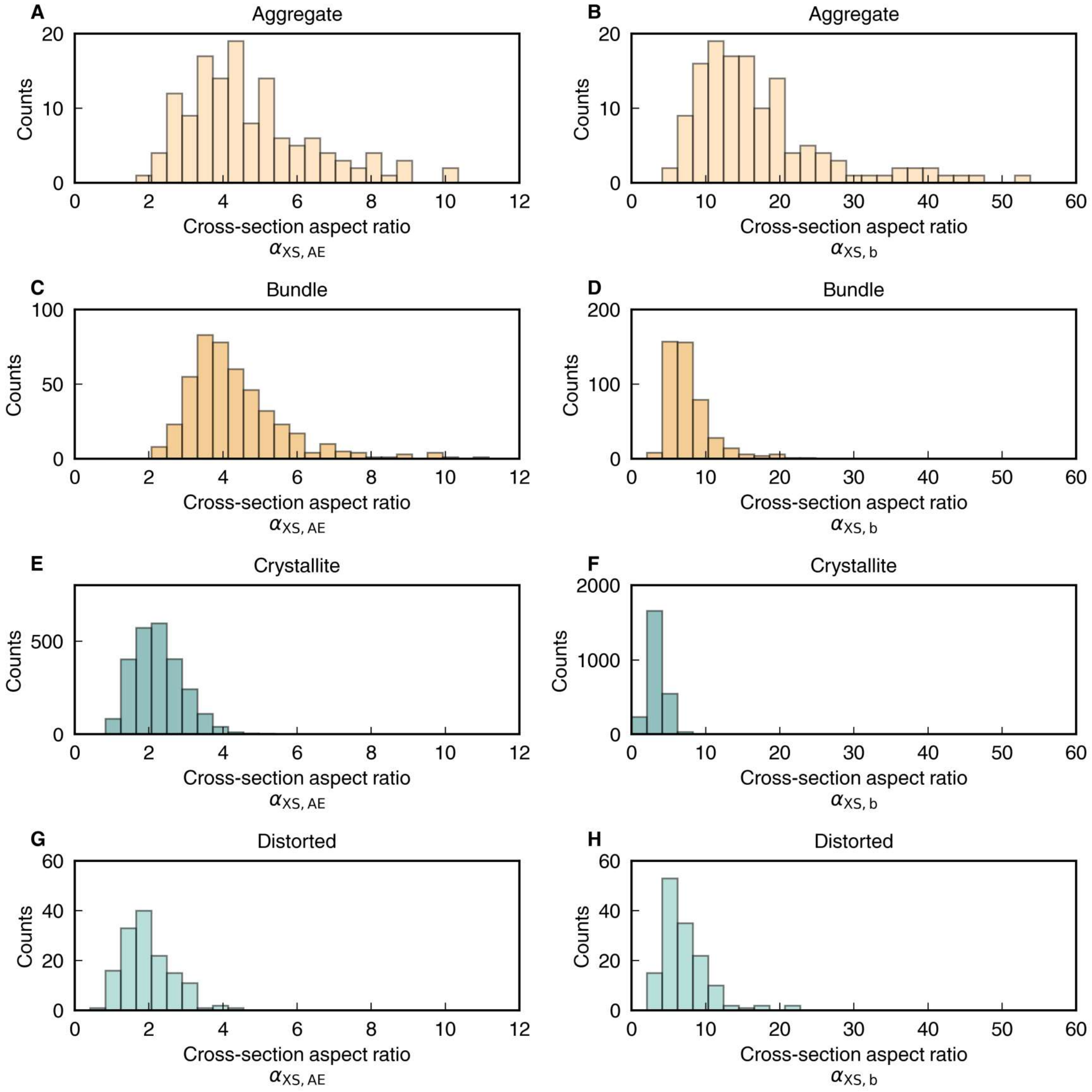

**Figure S19:** Histograms for $\alpha_{XS,AE}$ and $\alpha_{XS,b}$ for each particle class.

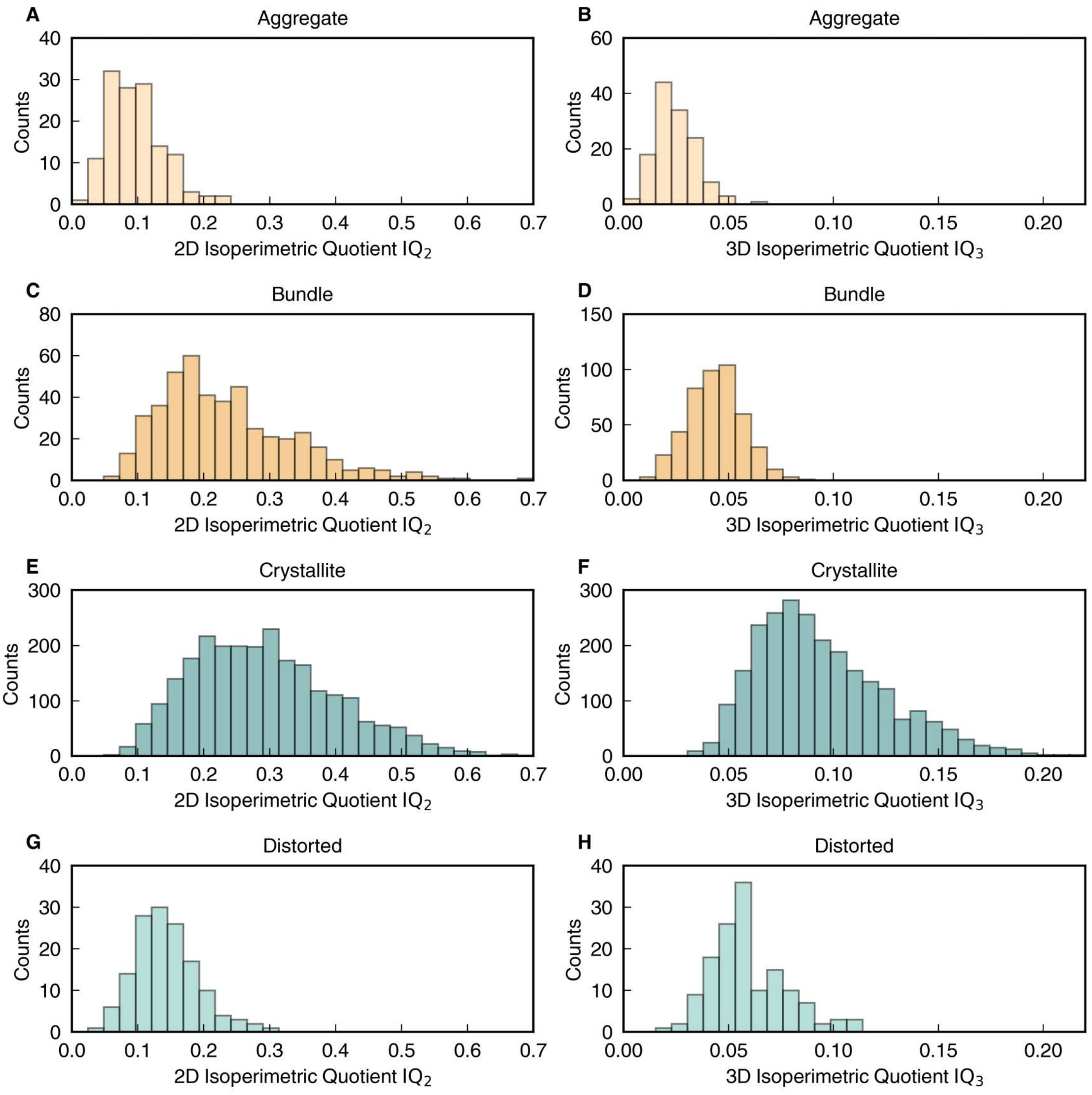

**Figure S20:** Histograms for $IQ_2$ and $IQ_3$ for each particle class.

# S12 Estimation of particle cross-section using turbidity spectra

Sonication visibly reduces the turbidity (*i.e.* cloudiness) of CNC suspensions due to the breakdown of scattering particles. The decrease in turbidity with sonication dose was determined by UV-vis transmission spectroscopy (Section S1.7) and is shown in Figure S21.

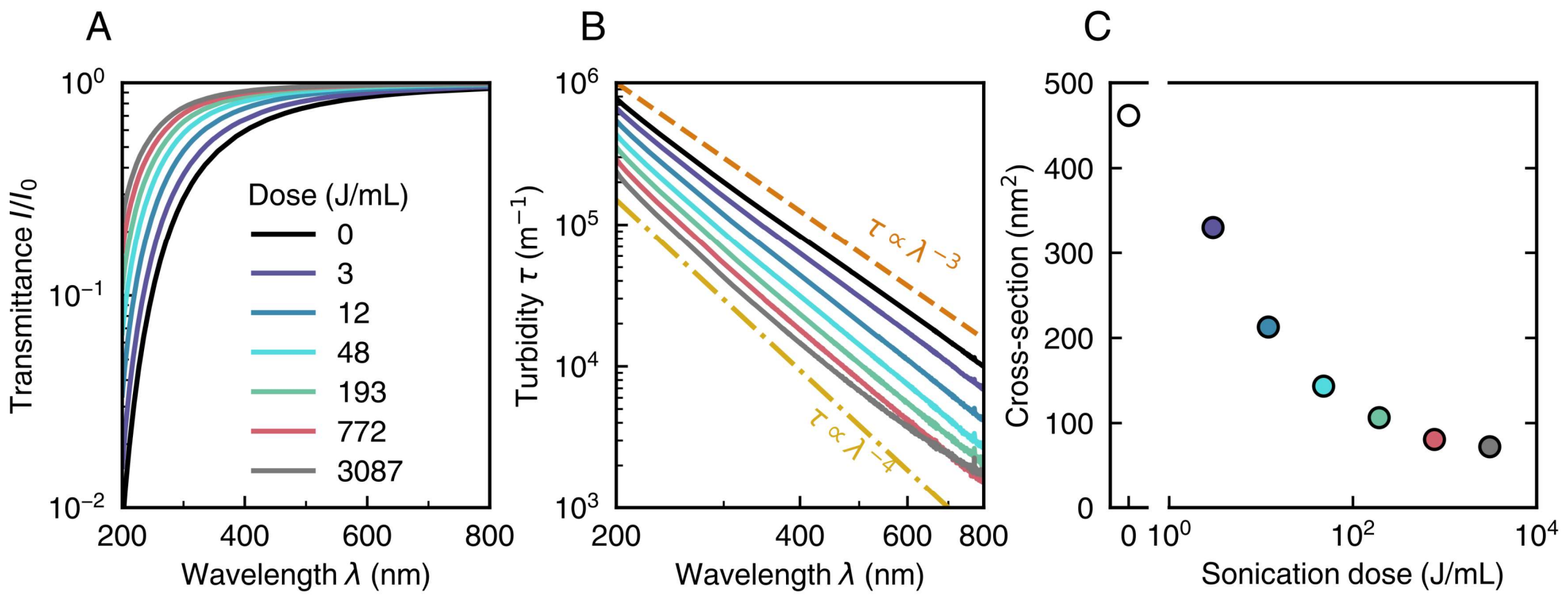

**Figure S21:** (**A**) Transmission through 0.1 wt% CNC suspensions increases with sonication dose (**B**) The power-law scaling of the turbidity spectra is clearer in a log-log plot (**C**) Volume-weighted mean particle cross-section versus sonication dose.

In a log-log plot of turbidity versus wavelength (Figure S21b), it is clear that alongside an overall decrease in turbidity at all wavelengths, sonication changes the shape of the turbidity spectrum, with a $\lambda^{-3}$ wavelength dependence at low sonication dose transforming into a $\lambda^{-4}$ dependence at high dose. This behaviour can be understood in terms of the morphological changes induced by sonication, and can be used to estimate the mean CNC particle cross-section (17).

For a turbid suspension of scattering dielectric particles in the limit of infinite dilution, light transmission obeys a Beer-Lambert-type decay with optical path length $z$:

$$I(z) = I_0 \exp(-kz). \tag{14}$$

For a suspension of identical particles $p$, the turbidity is given by $\tau = k/c_p$, where $k$ is the decay constant in Equation (14) above and $c_p$ is the particle concentration expressed as the number of particles per volume. If the particles are small compared to the wavelength ($v_p^{-1/3} < \lambda$) and have mild refractive index contrast with the solvent medium ($(n_p - n_0)/n_0 \ll 1$), the turbidity can be approximated using the Rayleigh-Gans-Debye model:

$$\tau_p = 24\pi^3 \frac{n_0^4 v_p^2}{\lambda^4} \left( \frac{n_p^2 - n_0^2}{n_p^2 + 2n_0^2} \right)^2 Q_p, \tag{15}$$

where $Q_p$ is a geometric factor that depends only on the particle shape.

The turbidity of CNC suspensions can be understood by comparing two ideal shapes: isotropic spheres and long slender rods. For spherical particles, $Q_{sph} = 1$ and the turbidity is

$$\tau_{sph}(\lambda) = 24\pi^3 \frac{n_0^4 v_p^2}{\lambda^4} \left( \frac{n_p^2 - n_0^2}{n_p^2 + 2n_0^2} \right)^2, \tag{16}$$

which is the conventional expression for Rayleigh scattering with $\lambda^{-4}$ wavelength dependence. For a slender rod with an extended length $L_p \gg \lambda$ but sub-wavelength cross-section, the geometric factor is (18; 19)

$$Q_{rod} = \frac{11}{20} \frac{\lambda}{n_0 L_p}. \tag{17}$$

and the turbidity is therefore

$$\tau_{rod}(\lambda) = \frac{66\pi^3}{5}\frac{n_0^3}{\lambda^3}\frac{v_p^2}{L_p}\left(\frac{n_p^2 - n_0^2}{n_p^2 + 2n_0^2}\right)^2, \tag{18}$$

which has a $\lambda^{-3}$ wavelength dependence. At low sonication dose, CNC particles behave like slender rods and therefore exhibit a $\lambda^{-3}$ wavelength dependence. As sonication breaks apart the CNCs, reducing the particle volume and aspect ratio, their morphology becomes less elongated and a $\lambda^{-4}$ wavelength dependence emerges (as seen in section S12B).

The turbidity equations given by Equation (15) are difficult to apply directly to CNC suspensions because the number density of CNCs ($c_p$) cannot be directly measured, and there is considerable polydispersity in CNC size and shape. Therefore, we consider CNC suspensions as a population of different particle species $p$, each with different particle volumes $v_p$ but identical refractive indices $n_p = n_1$, suspended in a medium of index $n_0$. In this case, the turbidity decay constant in eq. (14) generalises to

$$k = \sum_p \tau_p c_p = 24\pi^3 \frac{n_0^4}{\lambda^4}\left(\frac{n_1^2 - n_0^2}{n_1^2 + 2n_0^2}\right)^2 \sum_p Q_p v_p^2 c_p \tag{19}$$

where $\sum_p$ indicates summation over all particle species. Experimentally, the most accessible concentration metric is the overall particle volume fraction, given by

$$\phi = \sum_p v_p c_p = \sum \phi_p. \tag{20}$$

We therefore define the turbidity per volume fraction $\tau' = k/\phi$, or explicitly

$$\tau' = 24\pi^3 \frac{n_0^4}{\lambda^4}\left(\frac{n_1^2 - n_0^2}{n_1^2 + 2n_0^2}\right)^2 \frac{\sum_p Q_p v_p^2 c_p}{\sum_p v_p c_p} \tag{21}$$

For a distribution of slender rods (with $Q_p$ given by Equation (17) for all particle species), we find that

$$\tau' = \frac{66\pi^3}{5}\frac{n_0^3}{\lambda^3}\left(\frac{n_1^2 - n_0^2}{n_1^2 + 2n_0^2}\right)^2 \langle A \rangle \tag{22}$$

where $\langle A \rangle$ is the volume-weighted mean of the particle cross-section:

$$\langle A \rangle = \frac{\sum_p (v_p^2/L_p) c_p}{\sum_p v_p c_p} = \frac{\sum_p (v_p/L_p)\phi_p}{\sum_p \phi_p} \tag{23}$$

Equation (22) can therefore be used to estimate the particle cross-section from the experimental transmission spectra (Figure S21C), a technique that has previously been reported for CNC suspensions (17). Only the long-wavelength data ($\lambda > 400$ nm) were used, as the assumptions used to derive Equation (22) are only valid in the long-wavelength limit. The mean cross-section $\langle A \rangle$ decreases with sonication dose (Figure S21c) and approaches a limiting value of $\approx 70\,\text{nm}^2$ at high sonication dose.

## References

[1] E. J. Foster, *et al.*, *Chemical Society Reviews* **47**, 2609 (2018). Publisher: The Royal Society of Chemistry.

[2] D. Nečas, P. Klapetek, *Open Physics* **10**, 181 (2012). Publisher: De Gruyter Open Access.

[3] A. Brinkmann, *et al.*, *Langmuir* **32**, 6105 (2016).

[4] M. Kaushik, C. Fraschini, G. Chauve, J.-L. Putaux, A. Moores, *Transm. Electron Microsc. Theory Appl* (2015).

[5] P. Vanysek, *CRC hand book of chemistry and physics* pp. 5–92 (1993). Publisher: CRC press.

[6] S. Beck, J. Bouchard, R. Berry, *Biomacromolecules* **12**, 167 (2011).

[7] X. M. Dong, T. Kimura, J.-F. Revol, D. G. Gray, *Langmuir* **12**, 2076 (1996).

[8] R. M. Parker, *et al.*, *Advanced Materials* **30**, 1704477 (2017).

[9] G. Guidetti, H. Sun, A. Ivanova, B. Marelli, B. Frka-Petesic, *Advanced Sustainable Systems* **5**, 2000272 (2021). _eprint: https://onlinelibrary.wiley.com/doi/pdf/10.1002/adsu.202000272.

[10] T. Abitbol, D. Kam, Y. Levi-Kalisman, D. G. Gray, O. Shoseyov, *Langmuir* **34**, 3925 (2018).

[11] A. G. Dumanli, *et al.*, *ACS Applied Materials & Interfaces* **6**, 12302 (2014).

[12] M. M. Tirado, C. L. Martínez, J. G. de la Torre, *The Journal of Chemical Physics* **81**, 2047 (1984). Publisher: American Institute of Physics.

[13] G. Schumacher, T. G. M. v. d. Ven, *Journal of the Chemical Society, Faraday Transactions* **87**, 971 (1991). Publisher: Royal Society of Chemistry.

[14] L. Onsager, *Annals of the New York Academy of Sciences* **51**, 627 (1949).

[15] A. Stroobants, H. N. W. Lekkerkerker, T. Odijk, *Macromolecules* **19**, 2232 (1986).

[16] H. H. Wensink, G. J. Vroege, *The Journal of Chemical Physics* **119**, 6868 (2003).

[17] M. Shimizu, *et al.*, *Macromolecular Rapid Communications* **37**, 1581 (2016).

[18] E. F. Casassa, *The Journal of Chemical Physics* **23**, 596 (1955). ISBN: 0021-9606 Publisher: American Institute of Physics.

[19] M. E. Carr Jr, J. Hermans, *Macromolecules* **11**, 46 (1978). ISBN: 0024-9297 Publisher: ACS Publications.